\documentclass{elsarticle}
\usepackage[latin9]{inputenc}
\usepackage{array}
\usepackage{float}
\usepackage{url}
\usepackage{graphicx}

\makeatletter

\providecommand{\tabularnewline}{\\}
\floatstyle{ruled}
\newfloat{algorithm}{tbp}{loa}
\providecommand{\algorithmname}{Algorithm}
\floatname{algorithm}{\protect\algorithmname}

\journal{arXiv}

\usepackage{algorithm,algpseudocode}

\@ifundefined{showcaptionsetup}{}{%
 \PassOptionsToPackage{caption=false}{subfig}}
\usepackage{subfig}
\makeatother

\begin{document}

\begin{frontmatter}{}

\title{Empowering End-users\\
with Object-aware Processes}

\author[dbis]{Kevin Andrews\corref{corresponding}}

\cortext[corresponding]{Corresponding Author}

\ead{kevin.andrews@uni-ulm.de}

\author[dbis]{Manfred Reichert}

\address[dbis]{Institute of Databases and Information Systems, Ulm University,
Germany}
\begin{abstract}
Business process management systems from various vendors are used
by companies around the globe. Most of these systems allow for the
full or partial automation of business processes by ensuring that
tasks and data are presented to the right person at the right time
during process execution. However, almost all established BPMS employ
the activity-centric process support paradigm, in which the various
forms, i.e., the main way for users to input data into the process,
have to be created by hand. Furthermore, traditional activity-centric
process management systems are limited in their flexibility as all
possible execution variants have to be taken into account by the process
modeler. Therefore, large amounts of research have gone into developing
alternative process support paradigms, with a large focus on enabling
more flexibly executable processes. This article takes one of these
paradigms, object-aware process management, and presents the concepts
we developed while researching the possibility of bringing the power
and flexibility of non-activity-centric process support paradigms
to the people that matter: the end-users working with the processes.
The contribution of this article are the concepts, ideas, and lessons
learned during the development and evaluation of the PHILharmonicFlows
runtime user interface, which allows for the generation of an entire
user interface, complete with navigation and forms, based on an object-aware
process model. This novel approach allows for the generation of entire
information systems, complete with data storage, process logic, and
now fully functional user interfaces in a fully generic fashion from
data-centric object-aware process models. The extensive evaluation
performed with the help of an empirical study and a large scale real-world
deployment of the process engine and generic user interface, shows
that the concepts presented in this paper abstract enough of the challenging
conceptual underpinnings of data-centric process management paradigms
to make them accessible to end-users when executing processes that
require high degrees of flexibility.
\end{abstract}

\end{frontmatter}{}

\section{Introduction\label{sec:Introduction}}

The increasing popularity of process modeling notations, like Business
Process Model and Notation (BPMN) are making process documentation
commonplace in many companies. Process automation through the use
of business process management systems (BPMS), however, is not as
commonplace as most BPMS still lack flexibility support, instead favoring
rigid processes with high repetitiveness. This is mostly due to the
BPMS employing the activity-centric process support paradigm, which
can be used to create activities for users along predefined paths
determined by a process model. User interaction occurs in the context
of human tasks via predefined forms into which users can write data.
Adapting these forms for any reason, such as the addition of new fields,
of creating variants of the forms for presentation to persons different
levels of permission is often a cumbersome and error prone task. In
particular, many information systems, whether process-based or not,
require individual data elements to not only be held in e.g. a database
for persistence, but also handed through various levels of software
architecture and interfaces until they are actually displayed for
input or output in a form. Finally, while most process management
systems offer additional tools like form designers and scripting languages
for capturing form logic, these are external to the actual process
model itself, and therefore not transparent to the process execution
engine. This has numerous drawbacks as the logic and data handled
by the form input activities are essentially black boxes to the rest
of the process model, which, in turn, means that changes to the process
can not automatically change associated forms and vice versa.

Another important aspect to consider is the flexibility of the actual
activity execution path defined in the process model. In human-centric
processes, in which most activities are form-based, this constitutes
the order in which forms are displayed to the various process participants,
often based on their roles and process progression. In most process-based
systems, deviations to this order are only possible if the process
model allows for them, as so-called \emph{ad-hoc changes} to the process
execution are not supported by any major system. More flexible process
support paradigms, such as artifact-centric processes \citep{Cohn2009},
case handling \citep{VanderAalst2005}, adaptive case management \citep{swenson2011taming},
or object-aware processes \citep{kunzle2013phd} move the focus away
from the sequence of activities users have to complete, towards the
data of individual business cases. This is essential when wanting
to model business cases that are not ``rigid'' and require large
amounts of flexibility while still offering the benefit of steering
users through the use of a process model.

While a lot of research has been conducted on modeling and execution
of process models for these alternative process support paradigms,
little to no research has focused on actually making these paradigms
usable by end-users. Even the object-aware paradigm, which was developed
specifically for processes with a high degree of user interaction,
previously only offered a user interface for ``expert'' users. This
article attempts to close this gap by presenting concepts for a user
interface that can adapt to any object-aware process model by generating
interface elements from the process model at runtime. The goal was
to develop a user interface that automatically adapts to users and
the tasks they wish to perform. Additionally, concepts are presented
for hiding the complexity of the underlying process model from the
end-user. Finally, eliminating the need for a predefined user interface
significantly increases productivity and reduces turnaround times
for changes to processes \citep{Kolb2012}.

Section \ref{sec:Fundamentals} discusses the fundamentals of object-aware
process management. Section \ref{sec:Presenting-Object-aware-Processe},
the main contribution of this article, presents fundamental concepts
for generating an entire user interface from an object-aware process
model at runtime. Section \ref{sec:Evaluation} presents our extensive
empirical and real-world evaluation of the implementation of our developed
concepts, whereas Section \ref{sec:Related-Work} examines related
work. Finally, Section \ref{sec:Summary} provides a summary and an
outlook on future work.
\begin{figure}[H]
\centering{}\includegraphics[width=1\columnwidth]{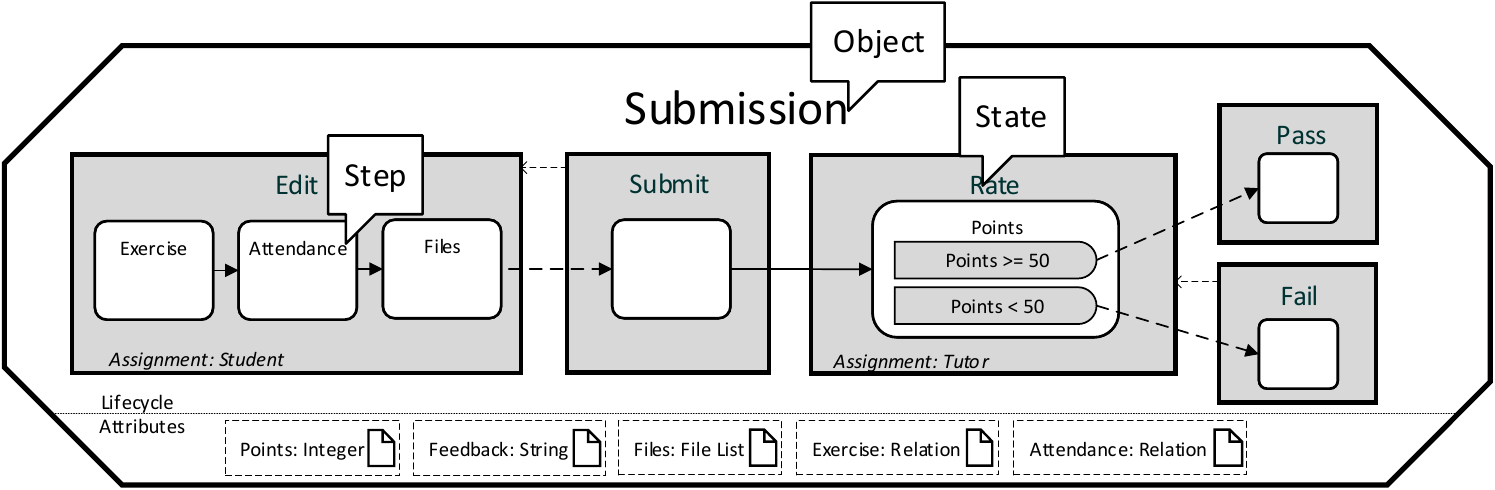}\caption{\label{fig:Example-PHILharmonicFlows-Object}Example Object Including
Lifecycle Process (Submission in PHoodle E-Learning)}
\end{figure}

\section{Fundamentals\label{sec:Fundamentals}}

This section provides an overview of the conceptual foundations of
object-aware process management, which are crucial for understanding
our work. PHILharmonicFlows, the object-aware process management framework
we are using as a test-bed for the concepts presented in this article,
has been under development at Ulm University \citep{steinau2018relational,kunzle2013phd,andrews2017enabling}
for many years. PHILharmonicFlows takes the idea of a data-driven
and data-centric process management system, enhancing it with the
concept of \emph{objects}. For each business object present in a real-world
business process one such object exists. As can be seen in Fig. \ref{fig:Example-PHILharmonicFlows-Object},
an object consists of data, in the form of \emph{attributes}, and
a state-based process model describing the data-driven \emph{object
lifecycle}. We utilize an object-aware process model of the PHoodle
(\textbf{PH}ILharmonicFlows M\textbf{oodle}) e-learning platform as
the running example for explaining the fundamentals of object-aware
process management in this article. The example model, when deployed
and executed, enables students to download exercise sheets and create
submissions for them while attending a university lecture. Furthermore,
other students working as tutors may rate their submissions.

\subsection*{Objects and Object Instances}

In object-aware process management, an \emph{object} only describes
the structure of its contained data and process logic at design-time
and an \emph{object instance} holds concrete data values and executes
the process logic at runtime. This may be compared to the concept
of a table and its rows in a relational database, with the table describing
the schema of the contained data and the rows containing the actual
data values at runtime. We further examine the concept of objects
utilizing the example of a \emph{Submission} object, which attendees
of a lecture may instantiate at runtime to hand in exercise or homework
submissions in our e-learning platform example.

\subsection*{Attributes and Lifecycle Processes}

The attributes of the \emph{Submission} object (cf. Fig. \ref{fig:Example-PHILharmonicFlows-Object})
include \emph{Points}, \emph{Feedback}, \emph{Files}, \emph{Exercise},
and \emph{Attendance}. The\emph{ lifecycle process}, in turn, describes
the different \emph{states} \emph{(Edit}, \emph{Submit}, \emph{Rate},
\emph{Pass}, and \emph{Fail)}, an instance of a \emph{Submission}
object may enter during process execution. Each state contains one
or more \emph{steps}, each referencing exactly one of the object attributes,
and enforces that the respective attribute is written at runtime.
The steps are connected by \emph{transitions}, which arrange them
in sequence. The state of the object changes when all steps of a state
are completed, i.e., after all attributes referenced by the steps
have been written. Furthermore, it is possible to assign \emph{permissions}
to attributes, even those that are not referenced by any steps in
the lifecycle process, e.g. \emph{Feedback}. These permissions can
be used to permit the reading or writing of attributes not being essential
to the execution of a state. Although largely omitted from this article,
permissions are essential in providing read access to attributes that
were supplied with data at earlier points in the object lifecycle
or write access to optional attributes, i.e., those that are not part
of the lifecycle process in the form of steps. Finally, alternative
and return paths through the lifecycle process are supported in terms
of \emph{decision steps} and \emph{backwards transitions}. An example
of a decision step is given by the \emph{Points} decision step in
the \emph{Rate} state, which branches the execution to the states
\emph{Pass} or \emph{Fail} based on the value of the \emph{Points}
attribute. On the other hand, an example of a backwards transition,
which allows the lifecycle process to return to an earlier state,
can be seen between the \emph{Submit} and \emph{Edit} states. To be
more precise, this backwards transition allows a \emph{Submission}
to be returned to the \emph{Edit} state while it is in the \emph{Submit}
state. However, as the model does not contain a backwards transition
from state \emph{Rate} to state \emph{Edit}, editing the \emph{Submission}
at a later point is impossible.

\subsection*{Lifecycle Process Execution}

As PHILharmonicFlows is \emph{data-driven}, the lifecycle process
execution of an instance of the \emph{Submission} object can be understood
as follows: The initial state of a \emph{Submission} is \emph{Edit}.
Once a \emph{Student} has entered data for attributes \emph{Exercise},\emph{
Attendance},\emph{ }and \emph{Files}, the \emph{Student} may trigger
the outgoing transition to the \emph{Submit} state (cf. Fig. \ref{fig:Example-PHILharmonicFlows-Object},
dashed line exiting step \emph{Files}). This causes the \emph{Submission}
to change its state to \emph{Submit}, in which it waits until the
submission period is over (cf. Paragraph \emph{Coordination}), after
which the Submission automatically transitions to state \emph{Rate}.
As state \emph{Rate} is assigned to a \emph{Tutor}, a user with role
\emph{Tutor} must input data for \emph{Points}. Based on the entered
value for \emph{Points}, the state of the \emph{Submission} object
either changes to \emph{Pass} or \emph{Fail}.

\subsection*{Form Generation}

\begin{figure}[H]
\centering{}\includegraphics[width=0.5\columnwidth]{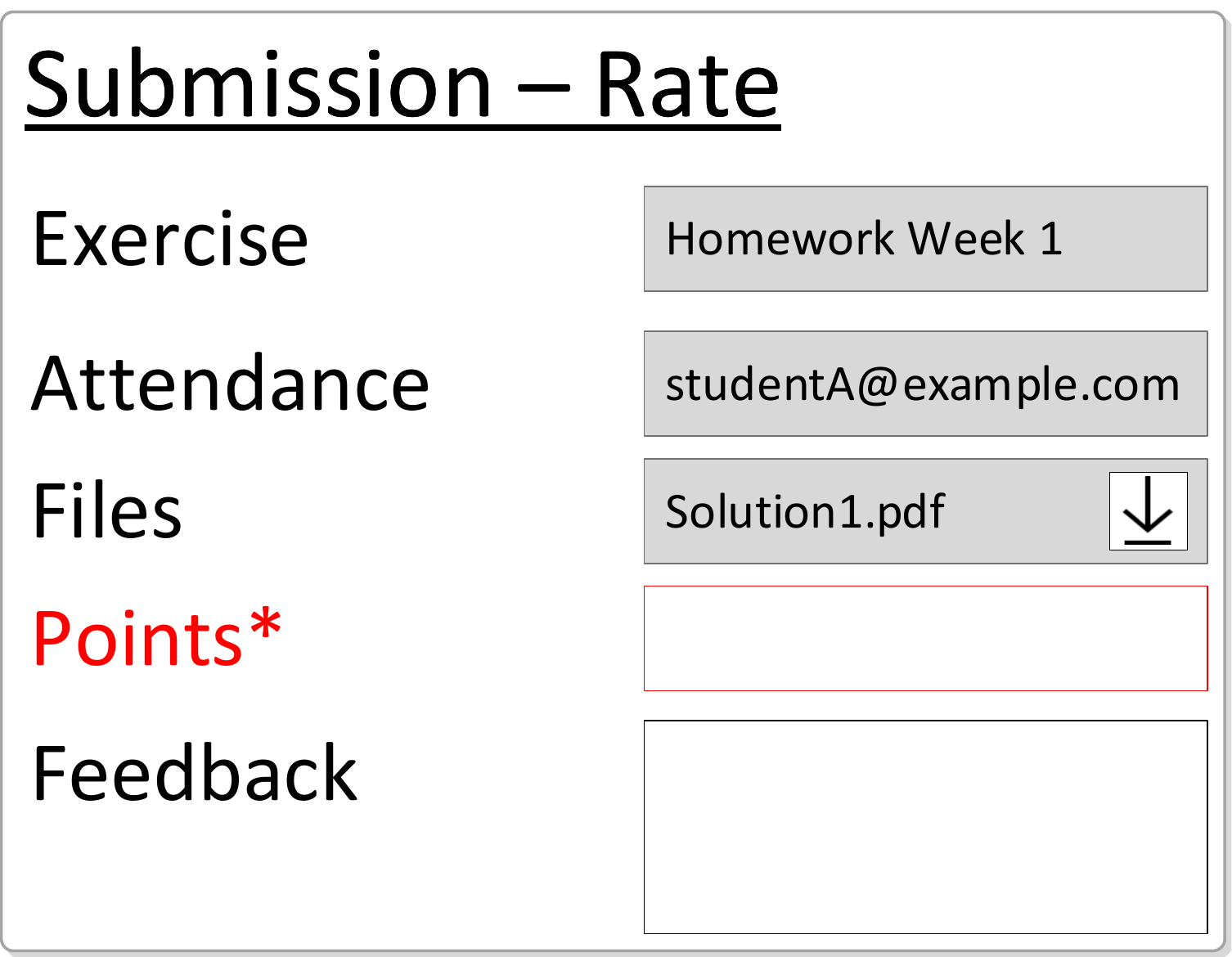}\caption{\label{fig:Example-PHILharmonicFlows-Form}Example Form (Tutor View
on a Submission)}
\end{figure}

Obviously, this fine-grained approach to modeling the individual parts
of a business process increases complexity compared to the activity-centric
paradigm, where the minimum granularity of a user action corresponds
to one atomic ``black box'' activity, instead of an individual data
attribute. However, as one of the major benefits, the object-aware
approach allows for \emph{automated form generation} at runtime. This
is facilitated by the lifecycle process of an object, which dictates
the attributes to be filled out before the object may switch to the
next state. This information is combined with the attribute read/write
permissions, resulting in a personalized and dynamically created form.
An example of such a form, derived from the lifecycle process in Fig.
\ref{fig:Example-PHILharmonicFlows-Object}, is shown in Fig. \ref{fig:Example-PHILharmonicFlows-Form}.

\subsection*{The Data Model and Data Model Instances}

\begin{figure}[H]
\centering{}\includegraphics[width=0.4\columnwidth]{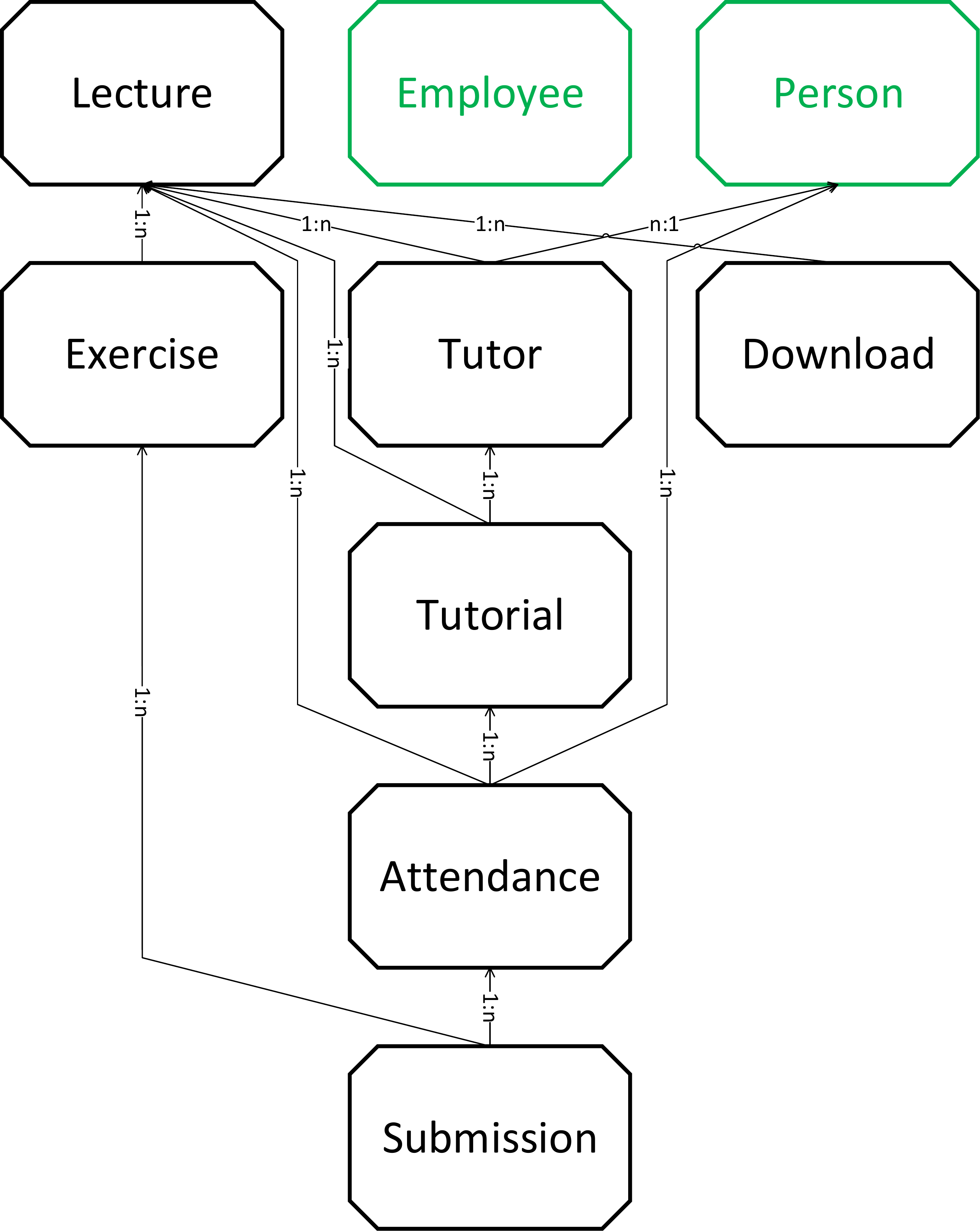}\caption{\label{fig:Design-time-data-model}Design-Time Data Model (PHoodle
E-Learning)}
\end{figure}

A single object and its resulting forms only constitute one part of
a complete business process in PHILharmonicFlows. To allow for more
complex executable business processes, many different objects and
users need to be involved \citep{steinau2018relational}. It is noteworthy
that \emph{users} are simply represented by special objects in the
object-aware process management concept. The entire set of objects
present in a PHILharmonicFlows process is denoted as the\emph{ data
model}, an example of which can be seen in Fig. \ref{fig:Design-time-data-model}.
The objects representing users are marked in green, i.e., there are
two different types of user objects in the e-learning data model,
\emph{Employees }(e.g. supervisors and staff) and \emph{Persons} (e.g.
students and tutors).

\begin{figure}[H]
\centering{}\includegraphics[width=0.4\columnwidth]{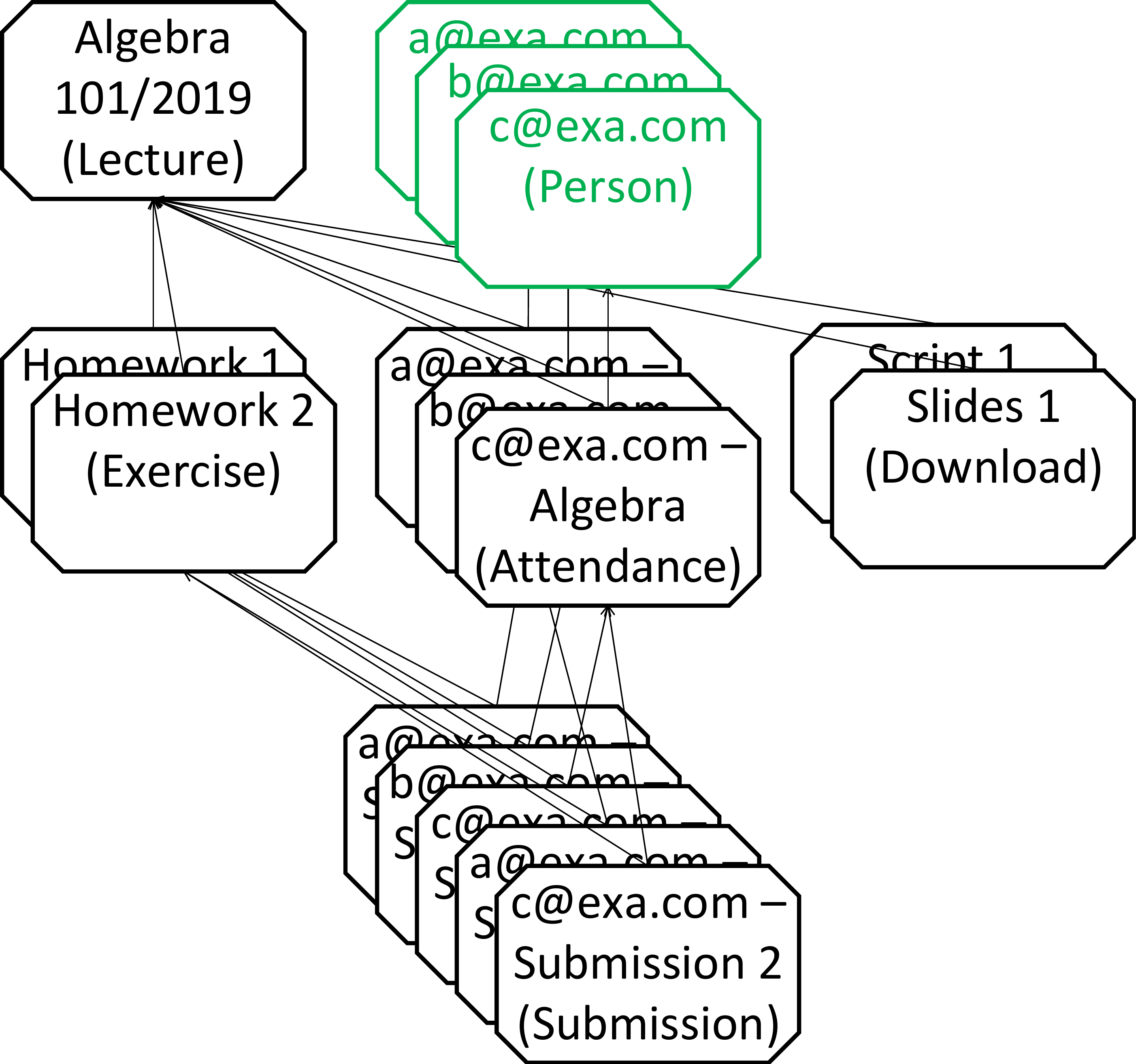}\caption{\label{fig:Run-time-data-structure}Runtime Data Model Instance (PHoodle
E-Learning)}
\end{figure}

In addition to the objects and users, the data model contains information
about the \emph{relations} existing between them. A relation constitutes
a logical association between two objects, e.g., a \emph{Submission}
and an \emph{Exercise}. At runtime, each of the objects may be instantiated
many times as \emph{object instances}.\emph{ }Note that the lifecycle
processes present in the various object instances may be executed
concurrently at runtime, thereby improving overall system performance.
Furthermore, the relations can also be instantiated at runtime, e.g.,
between an instance of a \emph{Submission} and an \emph{Exercise},
thereby associating the two object instances with each other. The
resulting meta information, expressing that the \emph{Submission}
in question belongs to the \emph{Exercise}, can be used to coordinate
the processing of the two object instances with each other at runtime
\citep{steinau2018relational}. Fig. \ref{fig:Run-time-data-structure}
shows an example of a \emph{data model instance} at runtime.

\subsection*{Coordination}

\begin{figure}[H]
\centering{}\includegraphics[width=1\columnwidth]{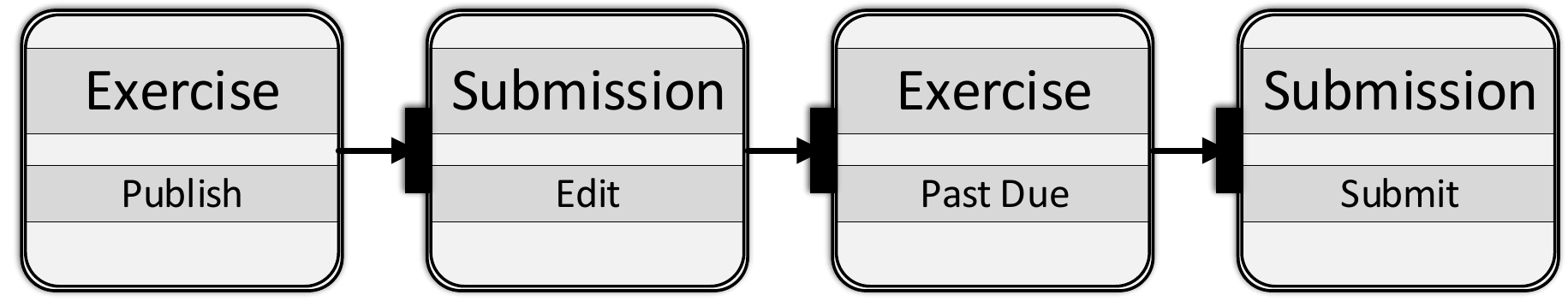}\caption{\label{fig:coordination}Coordination Example (Exercise vs. Submission)}
\end{figure}

The coordination of the object instances present in a data model instance
becomes necessary as business processes usually consist of hundreds
or thousands of interacting business objects \citep{muller2006flexibility},
whose concurrent processing needs to be synchronized at certain states.
As object instances publicly advertise their state information, the
current state of an object instance (e.g. \emph{Edit} or \emph{Submit})
can be utilized as an abstraction for coordinating its processing
(i.e., execution) with other object instances corresponding to the
same business process through a set of constraints and rules, defined
in a separate \emph{coordination process} \citep{steinau2018relational}.
As an example consider a set of constraints stating the following:
\begin{enumerate}
\item An \emph{Exercise} must be in state \emph{Publish} for \emph{Submissions}
that are related to it to be allowed to progress past state \emph{Edit}
(i.e., to state \emph{Submit}, cf. Fig. \ref{fig:Example-PHILharmonicFlows-Object}).
\item An \emph{Exercise} must be in state \emph{Past Due }for \emph{Submissions}
that are related to it to be allowed to progress past state \emph{Submit}
(i.e., to either state \emph{Pass} or \emph{Fail}, cf. Fig. \ref{fig:Example-PHILharmonicFlows-Object}).
\end{enumerate}
A simplified and abstracted coordination process representing these
constraints is shown in Fig. \ref{fig:coordination}.

\subsection*{Implementation Architecture}

In our current proof-of-concept prototype, the various higher level
conceptual elements of object-aware processes, i.e., objects, relations,
and coordination processes, are implemented as microservices. For
each object instance, relation instance, or coordination process instance,
one microservice instance is created at runtime, turning the implementation,
PHILharmonicFlows, into a distributed process management system for
object-aware processes. Each microservice only holds data representing
the attributes of its object. Furthermore, the microservice only executes
the lifecycle process of the object it is assigned to. The only information
visible outside the individual microservices is the current ``state''
of the object, which, in turn, is used by the microservice representing
the coordination process to properly coordinate the objects' interactions
with each other using the coordination process model (cf. Fig. \ref{fig:coordination}).

\section{Presenting Object-aware Processes to End-Users\label{sec:Presenting-Object-aware-Processe}}

After having discussed the fundamental concepts of object-aware processes
in Section \ref{sec:Fundamentals}, this section presents the core
concepts for making the execution of object-aware processes possible
for end-users. Our goal was not only to enable the execution of object-aware
processes for expert users (e.g. IT specialists), but also for end-users
having no knowledge of the process model, or even the concept of a
``process''. Ideally end-users should not even realize that they
are interacting with a process management system.

\begin{figure}[H]
\centering{}\includegraphics[width=0.7\columnwidth]{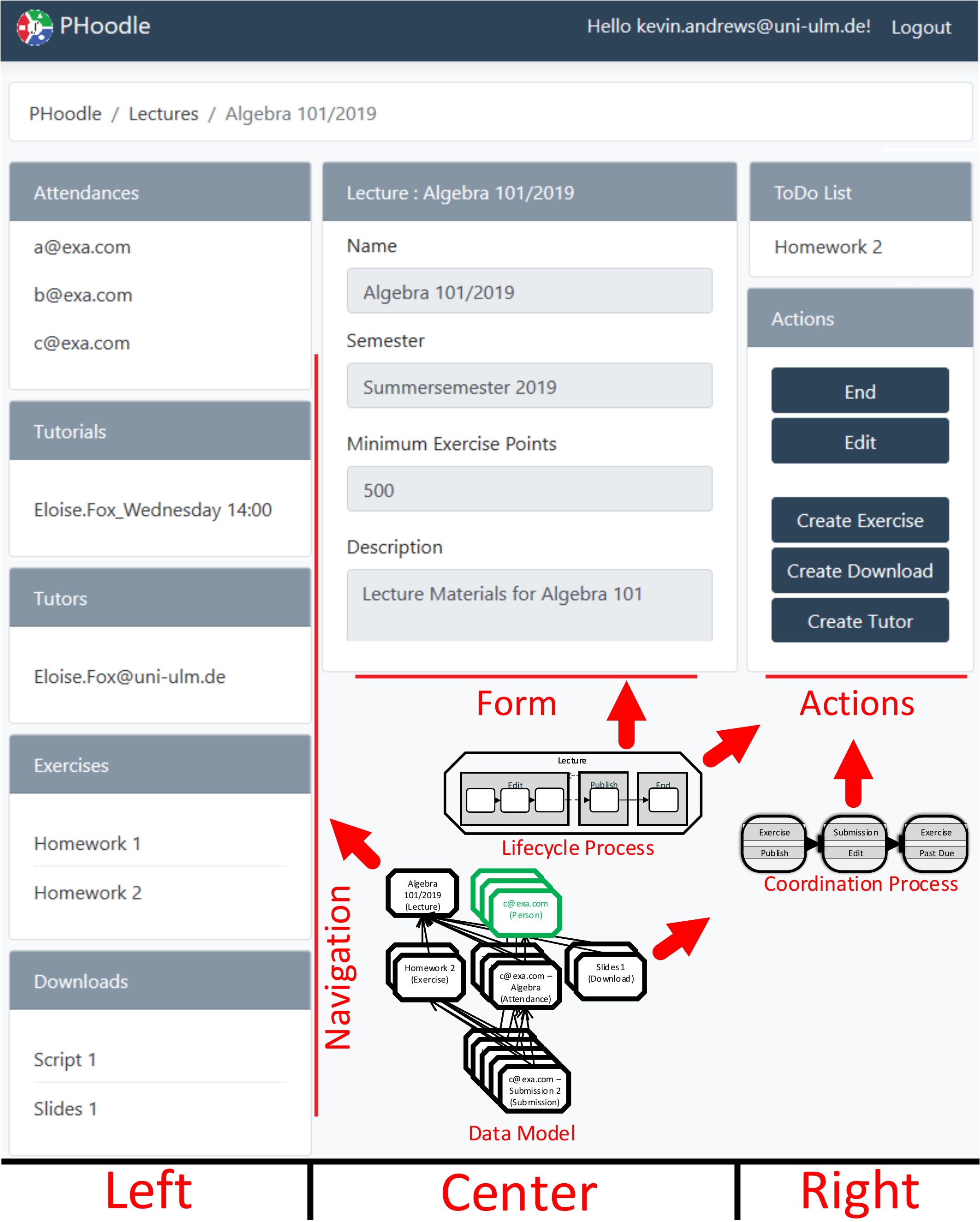}\caption{\label{fig:supervisoroverview}Supervisor View on Lecture ``Algebra''
(Desktop View)}
\end{figure}

To facilitate this, we developed a number of concepts for generating
entire user interfaces, including \emph{forms} and \emph{navigation}
elements, from processes created with the object-aware process management
paradigm. Furthermore, we implemented these concepts into the PHILharmonicFlows
engine and created a web-based runtime user interface for large scale
testing and verification of the correctness and usability of the concepts.
The result is a fully generic user interface that can load any object-aware
process and enable user interaction without any additional code. A
user interface prototype showing the supervisor view on a lecture
from our example data model instance (cf. Fig. \ref{fig:Run-time-data-structure})
is presented in Fig. \ref{fig:supervisoroverview}.

While this is the current version of the user interface, which we
are also employing for various studies and further research, keep
in mind that the goal of our research is to compress the complexity
of data-centric processes into a form that enables end-users to utilize
them and not to create the best possible website design. Note that,
apart from trivial functions such as ``Logout'', not a single string
in the user interface shown in Fig. \ref{fig:supervisoroverview}
is hard-coded, but instead derived directly from the conceptual process
model. Furthermore, no part of the user interface is hard-coded in
the process model either, i.e., there are no configuration files attached
to the process model determining which elements are placed where in
a form.

Examining the provided screenshot closer (cf. Fig. \ref{fig:supervisoroverview})
reveals three distinct vertical columns, starting below the breadcrumb-style
navigation helper. The left column gives an overview over objects
and their respective instances and offers users a way to \emph{navigate}
from object instance to object instance. The center column displays
a \emph{form} for the currently selected object instance. Finally,
the right column shows a general \emph{to-do list} as well as all
\emph{actions} currently available to the user. Note that the user
interface is always displayed for one specific user viewing one specific
object instance. A different user viewing the same lecture object
instance with a different role would see an entirely different page,
with different objects, form elements, and actions. As the generation
concepts for the content of the three columns are very different,
we dedicate the following three sections to one column each, starting
with the object instance navigation menu (left column).

\subsection{Generating the Navigation Menu\label{subsec:Generating-the-Navigation}}

While for very small examples it might be feasible to just display
a list of all object instances, perhaps grouped by objects, PHILharmonicFlows
supports very large data models with many different types of objects.
In this scenario, such a list would quickly cause a lack of overview
and confuse users as they constantly see objects not relevant to their
current work. Earlier iterations of the PHILharmonicFlows user interface
tackled this problem in a naive way by providing filtering tools,
allowing users to search for the type and identifier of the object
instance they wanted to interact with. However, this filter-based
navigation concept was completely replaced when developing the concepts
presented in this article. To be more precise, it was simply too cumbersome
and required knowledge from end-users about the data model itself
in order to ``know'' which objects they had to interact with, and
when.

Instead, the concept for navigating object instances presented in
this article utilizes the data model and the therein contained relations,
as well as the permission the user has, to only present the most relevant
object instances. In the following we illustrate basic concepts, along
the data model instance from the example presented in Fig. \ref{fig:Run-time-data-structure}.
The entirety of all object instances present in this data model instance
is a set of 1 \emph{Lecture}, 2 \emph{Downloads}, 3 \emph{Persons},
3 \emph{Attendances} (linking the Persons to the Lecture), 2 \emph{Exercises},
and 5 \emph{Submissions}.

\begin{figure}[H]
\centering{}\includegraphics[width=0.7\columnwidth]{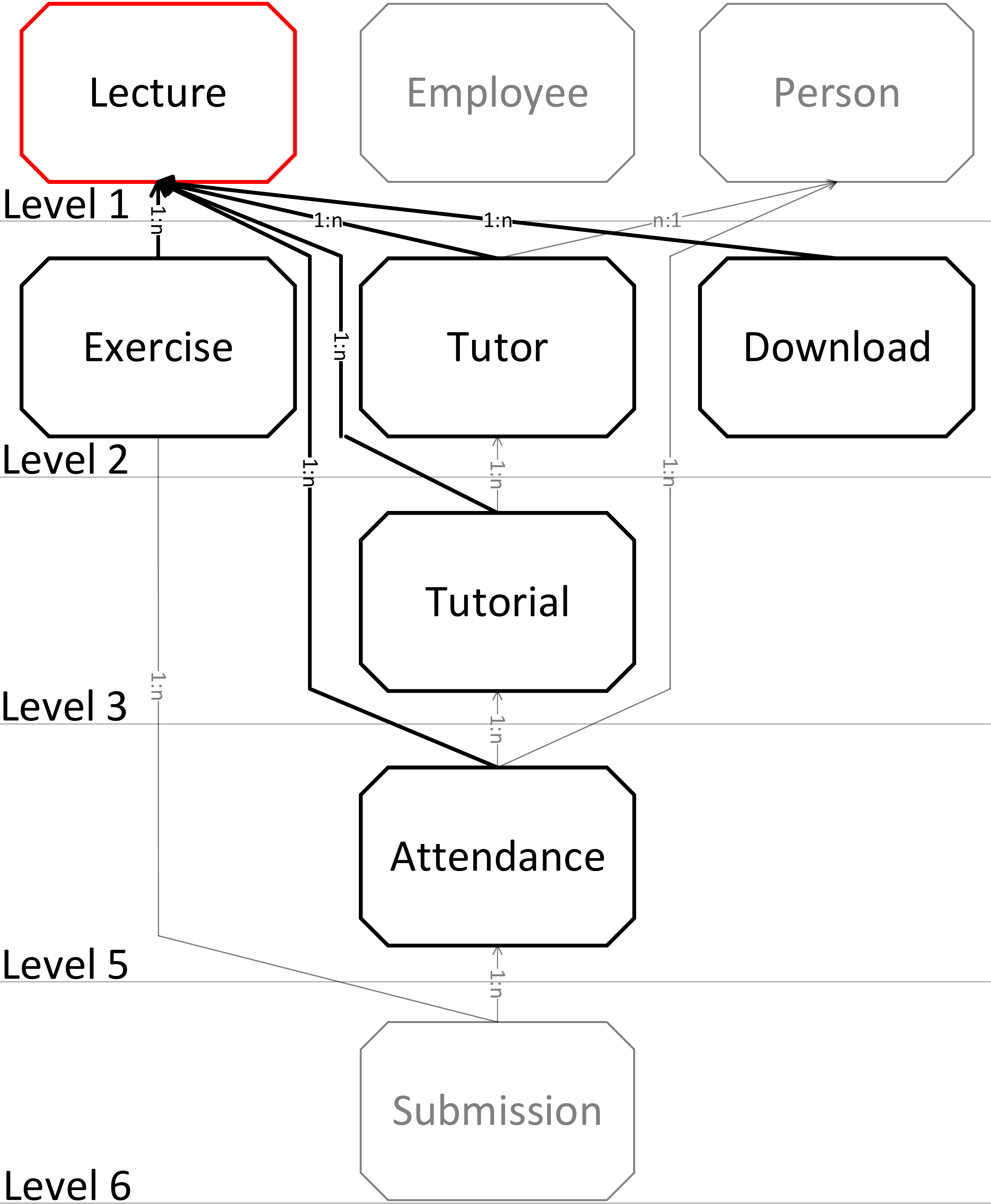}\caption{\label{fig:relationpaths}Objects with direct relations to \emph{Lecture}}
\end{figure}

First, when deciding which object instances shall be shown in the
navigation menu, it must be determined which object instance the user
is currently examining. Usually, this is simply the result of the
previous navigation command. Next, it must be determined to which
object instances the user most likely wants to navigate next. To this
end, we employ the relations that exist between the various object
instances. As the relations between the objects form a directed acyclic
graph at design-time, it is possible to organize the data model into
various levels, with the objects on lower levels ``belonging'' to
their respective higher level objects. In particular, the concept
only considers instances of objects from lower levels that are \textbf{directly}
related to the current object. Fig. \ref{fig:relationpaths} highlights
the relations from other objects to the \emph{Lecture} object that
constitute direct (i.e., non-transitive) relations.

Note that the information on which objects are directly related to
the current object can be analyzed statically based on the design-time
data model. Moreover, its offers an initial filter for the object
instances to be displayed. Furthermore, for an \emph{Employee} with
role \emph{Supervisor} (i.e., a user having the permissions to view
all object instances belonging to an object), this quick graph analysis
is sufficient to determine which object instances should be displayed
in the navigation menu. For example, this is reflected in the navigation
menu of the supervisor view on a lecture (cf. Fig. \ref{fig:supervisoroverview},
left column), which was generated based on the analysis of the data
model from Fig. \ref{fig:relationpaths}.

\begin{figure}[H]
\centering{}\includegraphics[width=0.4\columnwidth]{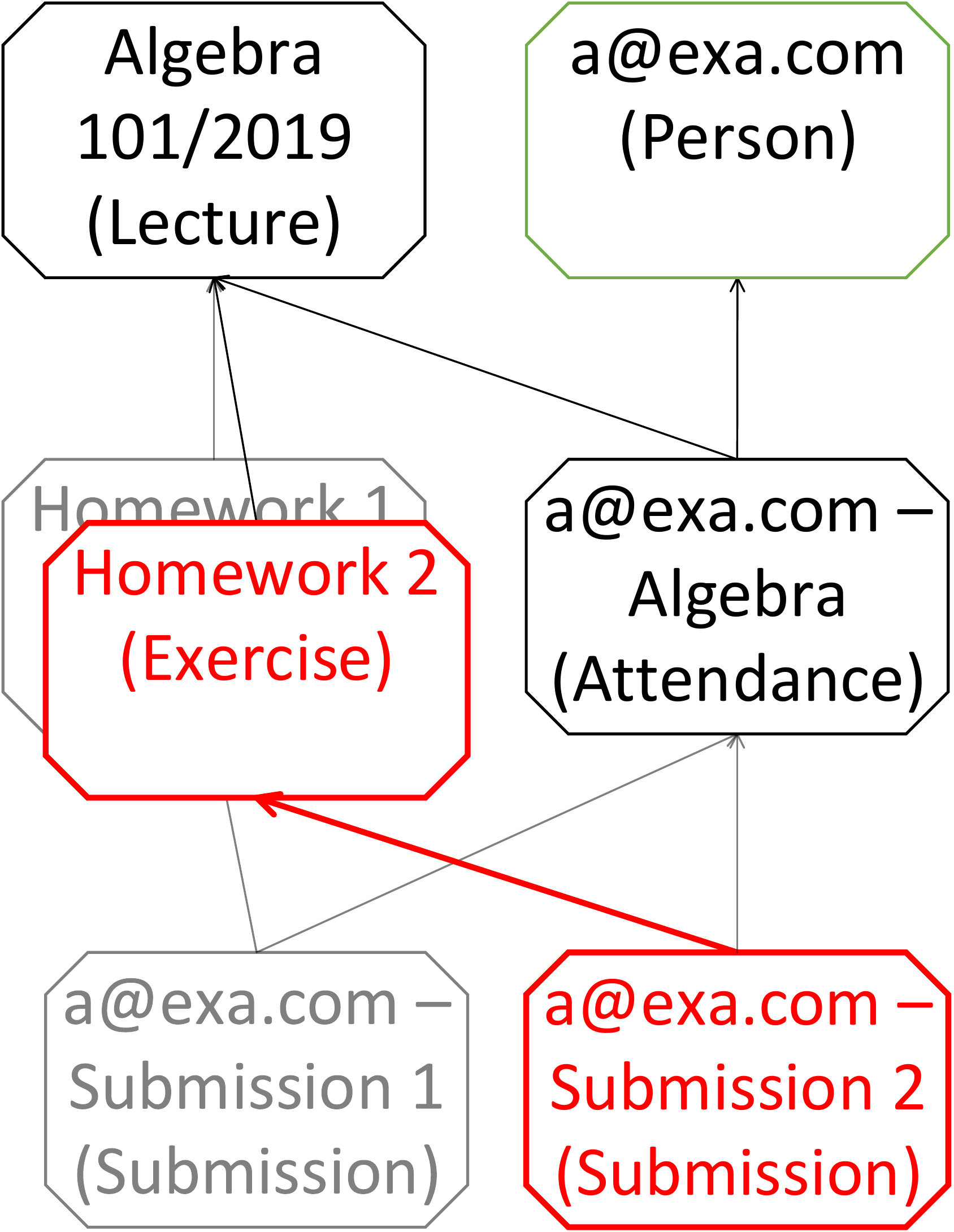}\caption{\label{fig:filteredoverexercise}\emph{Submissions }filtered via an
\emph{Exercise}}
\end{figure}

For users with restrictions on the object instances they may view,
such as \emph{Persons} with the \emph{Student} role, a more sophisticated
approach is required. In addition to the static analysis based on
the data model, filtering based on the type of object currently being
examined (i.e. the \emph{context} object instance) as well as the
permissions and actual relation instances existing at runtime have
to be taken into account. For example, a student not having any permissions
on the \emph{Tutor} object should not see any \emph{Tutor} object
instances. This is accomplished by the extensive permission system
employed by object-aware process management (cf. \citep{andrews2017enabling}
for details). Furthermore, and far more important regarding the contribution
of this article, it becomes necessary to analyze the relation instances
between the object instances at runtime, i.e., a dynamic analysis
of the data model instance is required while the navigation menu is
being generated. Fig. \ref{fig:filteredoverexercise} shows the part
of the data model instance that can be seen by the \emph{Person} \emph{a@exa.com
}after applying basic permission filtering, i.e., after removing all
object instances his role, i.e., \emph{Student}, does not have permissions
to view.

\begin{figure}[H]
\centering{}\includegraphics[width=0.6\columnwidth]{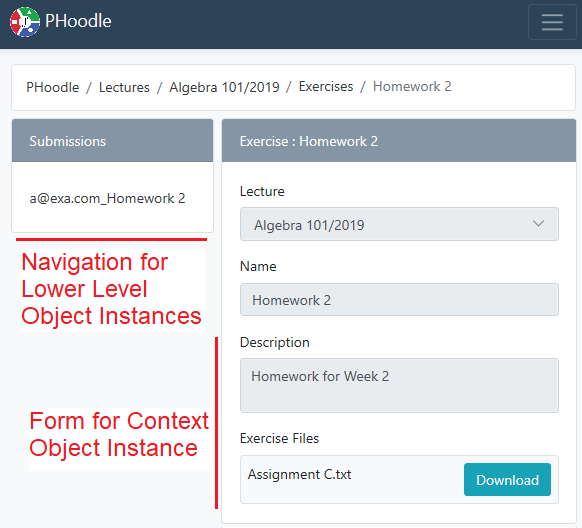}\caption{\label{fig:filteredoverexercisescreenshot}User Interface filtered
via an \emph{Exercise }(Cropped Tablet View)}
\end{figure}

Furthermore, note that \emph{Exercise} \emph{Homework 2 }is the \emph{context}
object instance for which the dynamic filtering should occur. The
context object instance determines the source parameter for an algorithm
that filters to lower level object instances with an existing relation
instance to the context object instance. In the example from Fig.
\ref{fig:filteredoverexercise}, this filters the lower level object
\emph{Submission} to the only instance having a relation instance
to the context object instance \emph{Homework 2}, i.e., \emph{Submission
2}. In our approach, this analysis results in the user interface shown
in Fig. \ref{fig:filteredoverexercisescreenshot} being generated
on the fly.

\begin{figure}[H]
\centering{}\includegraphics[width=0.4\columnwidth]{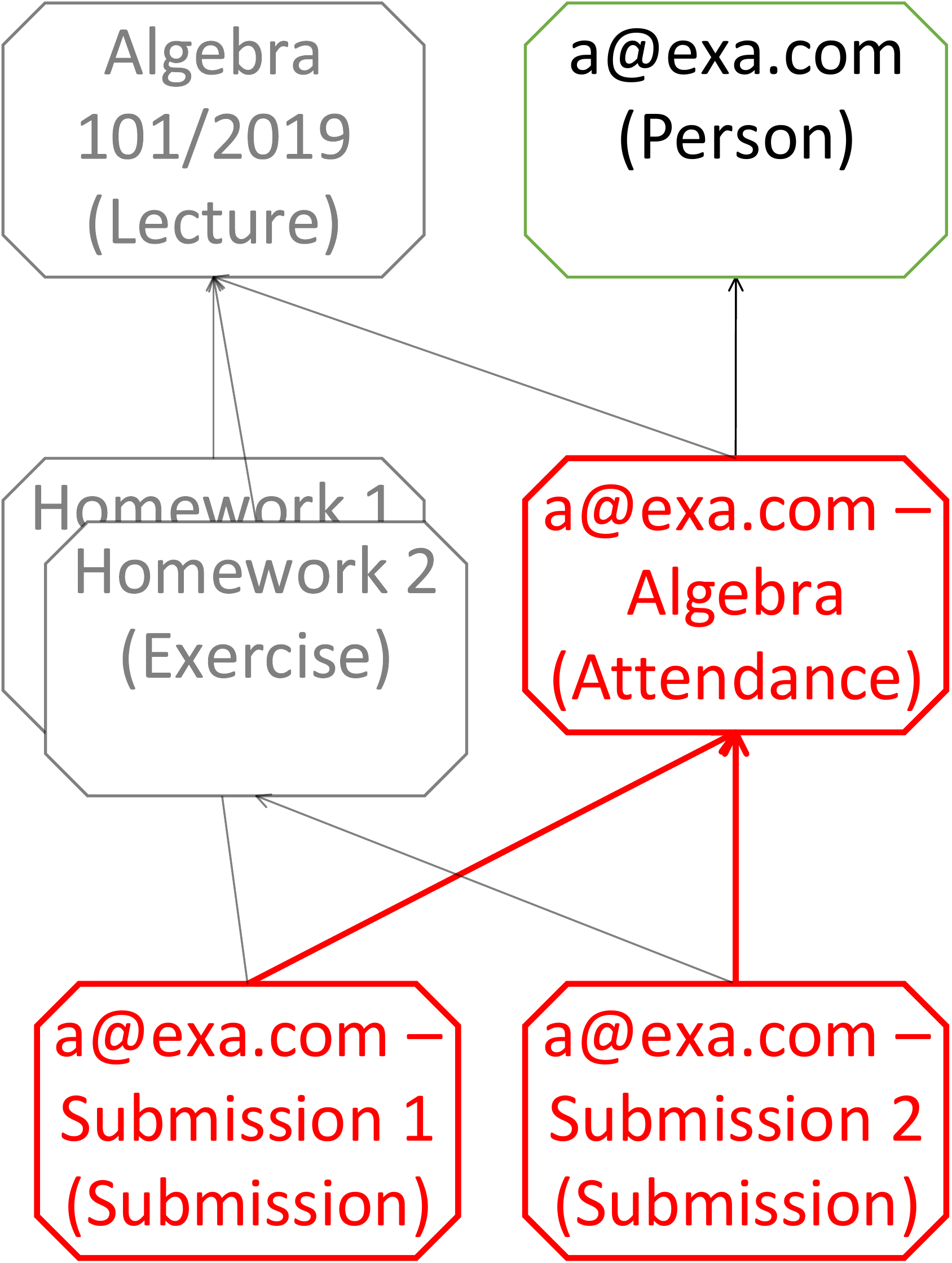}\caption{\label{fig:filteredoverattendance}\emph{Submissions }filtered via
an \emph{Attendance}}
\end{figure}

\begin{figure}[H]
\centering{}\includegraphics[width=0.6\columnwidth]{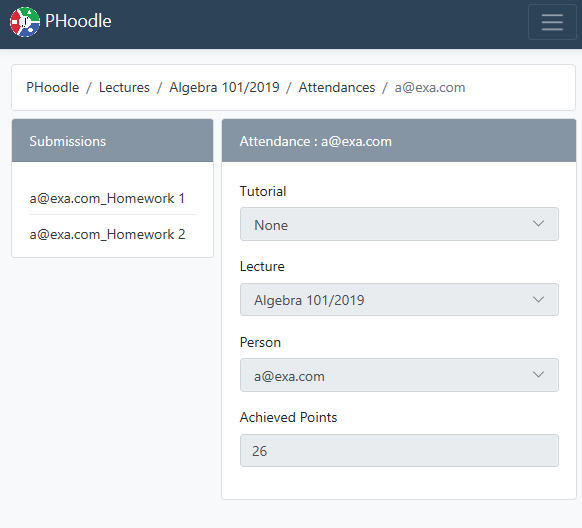}\caption{\label{fig:filteredoverattendancescreenshot}User Interface filtered
via an \emph{Attendance }(Cropped Tablet View)}
\end{figure}

Note that the navigation only consists of one possible object, as
for \emph{Homework 2}, \emph{Submission 2} is the only lower level
object instance related to \emph{Exercise} \emph{Homework 2}, which
is the current context object in Fig. \ref{fig:filteredoverexercisescreenshot}.
Conversely, if the user did not navigate to \emph{Homework 2}, but
instead to, for example, his own \emph{Attendance} object instance
for \emph{Lecture} \emph{Algebra}, the situation shown in Fig. \ref{fig:filteredoverattendance}
would occur.
Note that in Fig. \ref{fig:filteredoverattendance} the \emph{Attendance}
object \emph{a@exa.com-Algebra} is the context object instance of
the filtering operation which is used to generate the navigation menu.
In consequence, the lower level \emph{Submission} object is filtered
to those instances that have a direct relation to the context object
instance, i.e., \emph{Submission 1} and \emph{Submission 2}. As this
does no longer filter by \emph{Exercise} the resulting user interface
generated during the view creation shows navigation options for the
\emph{Submissions} of both \emph{Exercises} (cf. Fig. \ref{fig:filteredoverattendancescreenshot}).

While these examples have been fairly limited in scope, the concept
for generating the navigation menu from the relations of the current
context object instance, i.e., the object the user wishes to interact
with or show an input form for, works very well, even for large scale
data model instances. However, corner cases, in which a very large
number of object instances are related to the context object instance
and where the user also has permissions allowing access to all of
them, do exist. These can be handled in a traditional fashion, e.g.,
with a simple text-based filter on the navigation element that is
generated for the object in question, allowing quick filtering via
the label of the individual object instances.

\subsection{Generating the Forms\label{subsec:Generating-the-Form}}

After completing the navigation column, the form for interacting with
an object instance is generated. Automated form generation has been
a goal of the object-aware process management paradigm from early
on. Most lifecycle-based process management approaches support form
generation by analyzing the lifecycles present in the process model.
However, object-aware process management goes a step further by incorporating
a sophisticated permission system as well \citep{andrews2017enabling}.
In particular, the permission system implemented in the PHILharmonicFlows
engine allows for the definition of permissions based current attribute
values of on an object instance, as well as roles and permissions
which are dynamically evaluated based on the relations that exist
between objects at runtime.

\begin{figure}[H]
\centering{}\includegraphics[width=0.7\columnwidth]{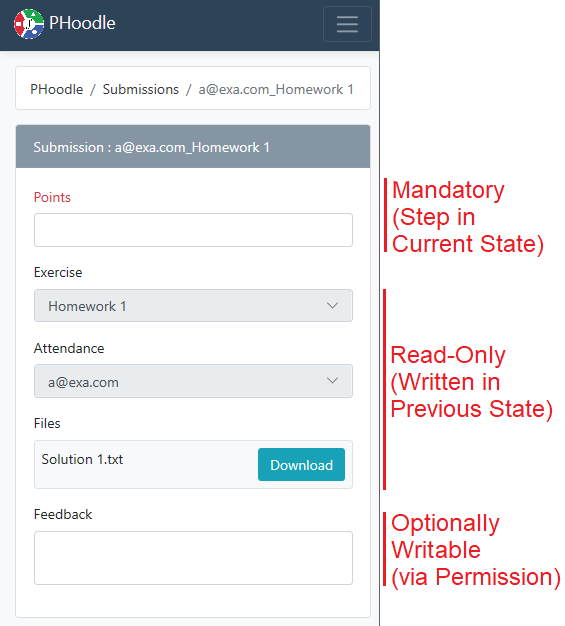}\caption{\label{fig:tutorformnopoints}Form allowing a \emph{Tutor} to \emph{Rate}
a \emph{Submission }(Mobile View)}
\end{figure}

The combination of the information gathered from the lifecycle process
of an object instance and the permissions the current user has on
the various attributes present in the current context object instance,
can be utilized for fine-grained form generation. On one hand, the
steps of the lifecycle process offer information on the mandatory
form fields the user has to provide values for in order to advance
the object instance to the next state. On the other, permissions are
utilized to allow reading or writing form fields for optional attributes,
which are not mandatory as part of the lifecycle execution or have
already been written beforehand. An example of such a form, in particular
a screenshot of the form concepted in Fig. \ref{fig:Example-PHILharmonicFlows-Form}
and derived from the \emph{Submission} object presented in Fig. \ref{fig:Example-PHILharmonicFlows-Object},
is shown in Fig. \ref{fig:tutorformnopoints}.
Note that there is no navigation column on the left as a \emph{Submission}
is on the lowest level of objects (cf. Fig. \ref{fig:relationpaths})
and there are no lower level object instances attached to it which
could be navigated to. Furthermore, there is no action column on the
right as there are no context actions for the current object instance
because the lifecycle process of \emph{Submission }objects requires
the \emph{Points} attribute to be written (i.e. provided a value)
before lifecycle process execution may continue to the next state
(cf. Fig. \ref{fig:Example-PHILharmonicFlows-Object}).

\subsection{Generating To-do Lists and Context Actions\label{subsec:Generating-To-do-Lists}}

\begin{figure}[H]
\centering{}\includegraphics[width=0.7\columnwidth]{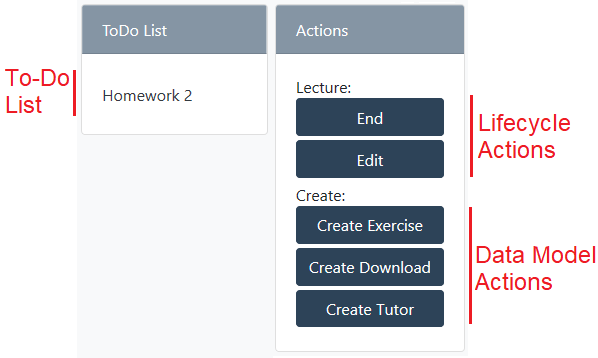}\caption{\label{fig:todoactions}To-do List and Actions Column from Fig. \ref{fig:supervisoroverview}
(Rearranged for Space)}
\end{figure}

This section explains the concepts behind the generation of the to-do
list and its corresponding actions available in the context of an
object instance. Fig. \ref{fig:todoactions} is a rearranged crop
of Fig. \ref{fig:supervisoroverview}, reiterating the parts of the
user interface relevant for the following sections.

\subsubsection{To-do List}

The \emph{to-do list} is actually fairly simple, as it is directly
derived from the role assignment of the current state of an object
instance. As discussed in Section \ref{sec:Fundamentals}, each object
lifecycle contains multiple states that may be traversed by corresponding
object instances during lifecycle execution. Each of these states
may have assigned roles, meaning that users possessing these roles
are supposed to advance the objects from the state in question to
the next state.

Regarding the to-do list element of the user interface, we combine
this assignment information with a relaxed version of the algorithm
for generating the navigation menu, which also takes into account
the object instances transitively related to the current context object
instance. This allows the user interface to display \textbf{any} lower
level object instance of the current context object instance in the
to-do list, as long as it is assigned to one of the roles the current
user has. For example the to-do list of an \emph{Employee} with role
\emph{Supervisor}, viewing a \emph{Lecture} that has relations to
instances of the \emph{Exercise} object, would show all \emph{Exercise}
object instances which are still in state \emph{Edit}, as is the case
for \emph{Homework 2} in Fig. \ref{fig:todoactions}. In our concept
evaluation study, the to-do list has proven to be an essential tool
in providing actual process support to users without knowledge of
the data model, guiding them to the next object instance with pending
work.

The ``action'' elements of the user interface are the actions available
in the context of an object instance. These can be categorized into
lifecycle actions and data model actions.

\subsubsection{Lifecycle Actions}

The \emph{lifecycle actions} available in the context of an object
instance correspond to the actions dealing with state traversal. This
includes advancing an object instance to the next state with a transition,
or returning to a previous state with a backwards transition (cf.
Section \ref{sec:Fundamentals}). In particular, early concepts of
the PHILharmonicFlows user interface relied on showing a model of
the lifecycle process to users, allowing them to click on transitions
to trigger state changes in a graph-based view. However, the idea
was discarded as it is not feasible to require users to have knowledge
of process models. Instead, the permission system was extended to
support transition permissions, thereby allowing the user interface
to present state traversal actions only to those users having the
permission to advance to a successor or predecessor state using a
transition. In consequence, there are lifecycle actions available
to an \emph{Employee} with the \emph{Supervisor} role in the context
of a \emph{Lecture}, namely \emph{End} and \emph{Edit} (cf. Fig. \ref{fig:todoactions}).
A user logged in as a \emph{Person} with the \emph{Student} or \emph{Tutor}
role, however, would not see these actions as the process model for
the e-learning example does not grant the necessary permissions.

On a side note, the labeling of actions \emph{End} and \emph{Edit}
is automatically derived from the lifecycle process model for the
\emph{Lecture} object, which was omitted from the article, as it only
contains three states labeled \emph{Edit}, \emph{Publish}, and \emph{End}.
As the context object for the user interface shown in Fig. \ref{fig:todoactions}
corresponds to a \emph{Lecture} in the \emph{Publish} state (cf. also
Fig. \ref{fig:supervisoroverview}), the transition from \emph{Publish}
to \emph{End} and the backwards transition from \emph{Publish} to
\emph{Edit} cause the generation of the corresponding buttons as actions
in the user interface.

\subsubsection{Data Model Actions\label{subsec:Data-Model-Actions}}

A fundamental concept added to object-aware process management consists
of the \emph{data model actions} offered to users in the PHILharmonicFlows
user interface. In particular, while a large part of this article
deals with the object instances and relation instances comprising
a data model instance, it has not been discussed how these relation
instances are created at runtime. Moreover, while concepts such as
relations between data as well as treating data as first-class citizens
in a process model, exist in many process support paradigms, the complexity
these concepts bring with them is largely ignored in existing works.
Note that for any process support paradigm to be usable in the real
world, the complexity of creating large data structures has to be
hidden from users.

To this end, we developed multiple extensions to the object-aware
process support paradigm. As a first extension, we introduced the
ability to define \emph{relation attributes}, i.e., object attributes
that hold a related object instance as their value. Similarly to restricting
regular attributes to common data types, such as String or Integer,
relation attributes are typed by the related object at design-time.
The \emph{Submission} object shown in Fig. \ref{fig:Example-PHILharmonicFlows-Object},
for example, contains two relation attributes that are typed to accept
instances of objects \emph{Exercise} and \emph{Attendance}, respectively.
Their representation as a drop down selector in the user interface
can be seen in Fig \ref{fig:tutorformnopoints} (albeit in their read-only
state).

In particular, this user interface element allows end-users to create
relations from one object instance to another simply by selecting
its name from a drop down list at runtime is a large step forward.
Moreover, allowing for the creation of relations to be, effectively,
forced as part of the object lifecycle removes ambiguities concerning
when and how objects must be related to each other. This allows process
modelers to capture rules such as ``a submission must always belong
to exactly one exercise and one attendee'' by simply forcing users
to set the corresponding relation attributes at runtime in the first
state of the \emph{Submission} lifecycle process. However, it is still
cumbersome to manually attach each newly created object instance to
the object instances it must be related to. Additionally, one must
explain the concept of relations to users, which is impractical in
real-world scenarios.

\begin{figure}[H]
\centering{}\includegraphics[width=0.9\columnwidth]{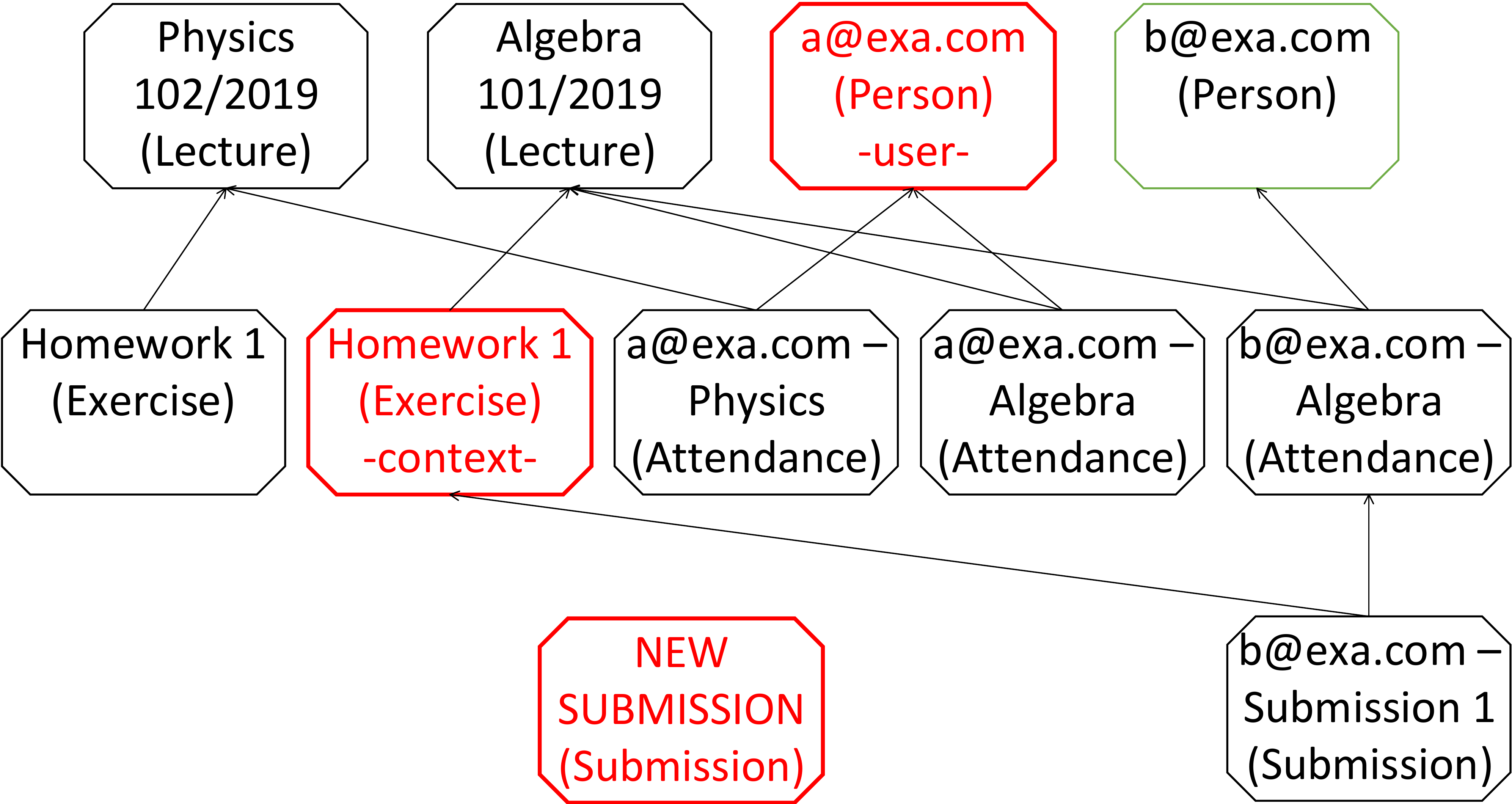}\caption{\label{fig:instandlinkimage}Example Data Model Instance for CORE
Algorithm}
\end{figure}

To alleviate these issues, we developed the CORE (Create Object Relations
Efficiently) algorithm (cf. Alg. \ref{alg:InstAndLink}). It takes
information present in an object-aware data model and combines it
with the newly added concept of relation attributes, thereby enabling
fully automatic relation creation when objects are instantiated by
users. To aid in understanding the CORE algorithm, Fig. \ref{fig:instandlinkimage}
gives an example of a typical situation in which the algorithm may
be applied.

Recalling that instances of the \emph{Person} object are always related
to a \emph{Lecture} via an \emph{Attendance} (enabling a user to attend
multiple lectures), it becomes obvious that \emph{Submissions} for
an \emph{Exercise} must always be related to (a) the \emph{Exercise}
and (b) the \emph{Attendance} corresponding to the \emph{Person} and
\emph{Lecture}. However, as the goal is to hide all this complexity
from users, instantiating a \emph{Submission} creates these relations
automatically. To facilitate this, the CORE algorithm (cf. Alg. \ref{alg:InstAndLink})
receives three pieces of information from the user interface: the
user currently logged in (\emph{Person a@exa.com}), the context object
instance, i.e., the object instance currently displayed in the user
interface (\emph{Exercise Algebra Homework 1}), and the object the
user wants to instantiate (\emph{Submission}).
\begin{algorithm}[H]
{\footnotesize{}\begin{algorithmic}[1]
\Require{$object$:Object, $context$:ObjectInstance, $user$:ObjectInstance}
\State $newObjectInstance \leftarrow object$.createInstance()
\State $firstState \leftarrow newObjectInstance$.lifecycle.states[0]
\ForAll{$step$ in $firstState$.steps()}
		\If{$step$.attribute \textbf{is} RelationAttribute}
			\If{$context$ \textbf{is} $step$.attribute.targetObjectType}
						\State $newObjectInstance$.createRelationTo($context$);
						\State \textbf{continue};
			\EndIf
			\ForAll{$userLowerLevelObjectInstance$ in $user$.lowerLevelObjectInstances}
				\If{$userLowerLevelObjectInstance$ \textbf{is} $step$.attribute.targetObjectType}
					\If{$context$.higherLevelObjectInstances $\cap$  $userLowerLevelObjectInstance$.higherLevelObjectInstances \textbf{==} 1}
						\State $newObjectInstance$.createRelationTo($userLowerLevelObjectInstance$)
					\EndIf
				\EndIf
			\EndFor	
		\EndIf
\EndFor
\end{algorithmic}}\caption{\label{alg:InstAndLink}Create Object Relations Efficiently (CORE)}
\end{algorithm}
When executing, the algorithm first creates an instance of the requested
object. Then, the algorithm loops over all relation attributes referenced
by steps in the first state of the lifecycle of the new \emph{Submission}
object instance (cf. Alg. \ref{alg:InstAndLink}, Lines 2-3). When
it encounters a relation attribute typed to the context object (\emph{Exercise}),
it creates a relation between the new \emph{Submission} and the \emph{Homework
1} context object instance (cf. Alg. \ref{alg:InstAndLink}, Lines
5-8). Furthermore, when it finds relation attributes typed to other
objects, e.g. \emph{Attendance}, it searches for instances of that
object which are related a) to the user object instance (\emph{Person
a@exa.com}) and b) to the context object instance (\emph{Exercise
Algebra Homework 1}). This is accomplished directly on the data model
instance graph by intersecting common higher level object instances
of a candidate object instance and the context object instance.

Regarding the example from Fig. \ref{fig:instandlinkimage}, the algorithm
would compile a list of all lower level \emph{Attendance} objects
of \emph{Person a@exa.com}, i.e., \emph{Attendance a@exa.com - Physics}
and\emph{ Attendance a@exa.com - Algebra}. This list would then be
filtered to solely include object instances having common higher level
instances with the context object instance. As the only higher level
object of the context object instance is \emph{Lecture} \emph{Algebra},
and only the \emph{Attendance a@exa.com - Algebra }shares this common
higher level object instance, the \emph{Algebra Attendance} is selected
as the target of a new relation, satisfying the \emph{Attendance}
attribute step in the lifecycle of the new \emph{Submission} object
instance (cf. Fig. \ref{fig:Example-PHILharmonicFlows-Object}) and
completing the CORE algorithm. In consequence, as the algorithm runs
during the instantiation of an object instance, the single click by
a user to create e.g. a \emph{Submission}, instantiates the object
and two relations instantly, completely hiding the complexity of the
relation concept from the user.

\section{Evaluation\label{sec:Evaluation}}

In order to study whether the presented user interface concepts make
object-aware processes usable in the real world, we conducted two
distinct evaluations: an empirical usability study utilizing multiple
scientific methods (cf. Section \ref{subsec:Empirical-User-Experience}),
and a large scale real-world deployment (cf. Section \ref{subsec:Real-World-Deployment-Measuremen})
of the PHoodle e-learning platform process model.

\subsection{Empirical User Experience Study\label{subsec:Empirical-User-Experience}}

This evaluation is based on a study conducted with n=70 subjects in
which multiple scenarios with various tasks had to be completed using
the PHoodle data model running on the PHILharmonicFlows process engine.
Interaction with the process engine was enabled by means of the fully
generic user interface prototype presented in this article. The study
utilized various established scientific methods for quantifying usability
and user experience, namely the After Scenario Questionnaire (ASQ),
the User Experience Questionnaire (UEQ), and a questionnaire according
to ISO norm 9241/110. The decision to employ all three methods simultaneously
was made to ensure that as much data as possible was gathered, as
each method focuses on slightly different aspects of usability.

\subsubsection{Subject Structure}

\begin{figure}[H]
\centering{}\includegraphics[width=0.6\columnwidth]{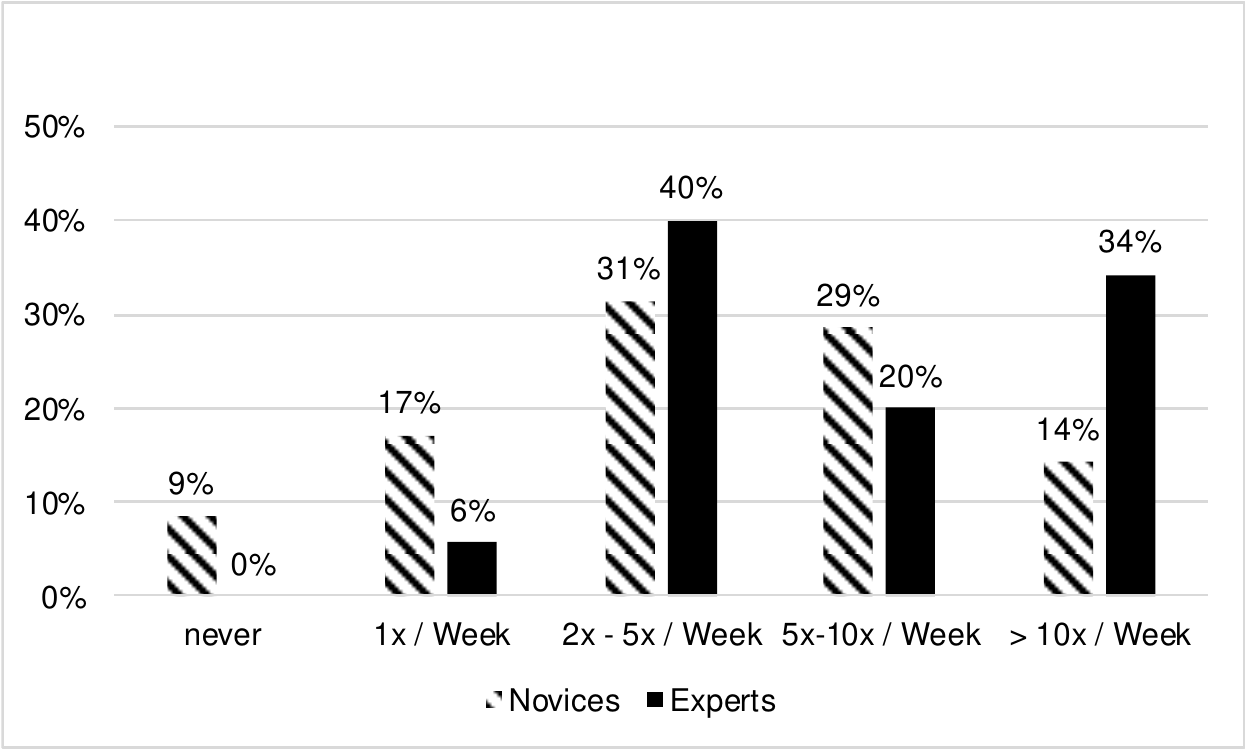}\caption{\label{fig:Usages-of-E-Learning}Usages of E-Learning Systems per
Week}
\end{figure}

The subjects participating in the study were divided into two groups,
consisting of 35 subjects each. The two groups, henceforth called
``novices'' and ``experts'', consist of employees and students
of Ulm University. The first group, the ``novices'' consists of
35 persons who had never worked with PHILharmonicFlows or the PHoodle
process model in any capacity. They knew nothing about the theoretical
concepts, the user interface implementation, the structure of the
data model or therein contained objects. While most of the novices
had used other e-learning systems or process management software prior
to the study, the novices had no experience or help in using PHILharmonicFlows
or the PHoodle process model. Note that almost 10\% of novice participants
had never used an e-learning system before (cf. Fig. \ref{fig:Usages-of-E-Learning}).

The ``expert'' group on the other hand consists of 35 students who
had been using PHoodle for two months prior to the study. They had
used PHoodle to submit homework and exercise sheets every week over
the course of these two months, i.e., each expert had completed at
least eight exercise submissions at the time the study was conducted.
Furthermore, they were given information on the usage of the PHILharmonicFlows
user-interface, as well as some background information on the underlying
concepts. Finally, the expert group also had a very high percentage
of subjects with very extensive and regular e-learning experience,
with 34\% percent of subjects reporting usage of over 10 times per
week and 94\% reporting at least two uses per week (cf. Fig. \ref{fig:Usages-of-E-Learning}).

\subsubsection{Scenarios}

Both groups, i.e., novices and experts, were asked to complete the
same five scenarios over the course of the study. The scenarios were
conducted on a demonstration system running the current PHILharmonicFlows
engine and hosting a web version of the user interface prototype presented
in this article. The data model was identical to the one used in our
real-world deployment and coincides with the data model presented
in the fundamentals (cf. Section \ref{sec:Fundamentals}). The data
model instance, however, was decoupled from the live data and reset
to a predefined state of object instances after each subject to ensure
that the actions of one subject could not influence the results of
another. The following paragraphs give a short overview over the five
scenarios.

\paragraph{Scenario 1 (Registration)}

The subjects are asked to register in the PHoodle user-interface with
a special e-mail address containing their assigned subject-specific
code, allowing them to be tracked across scenarios and correlate their
system actions with their questionnaire answers. The registration
is, from a user perspective, a very simple task. Accounts are unlocked
without e-mail confirmation for purposes of the study.

\begin{figure}[H]
\centering{}\includegraphics[width=0.9\columnwidth]{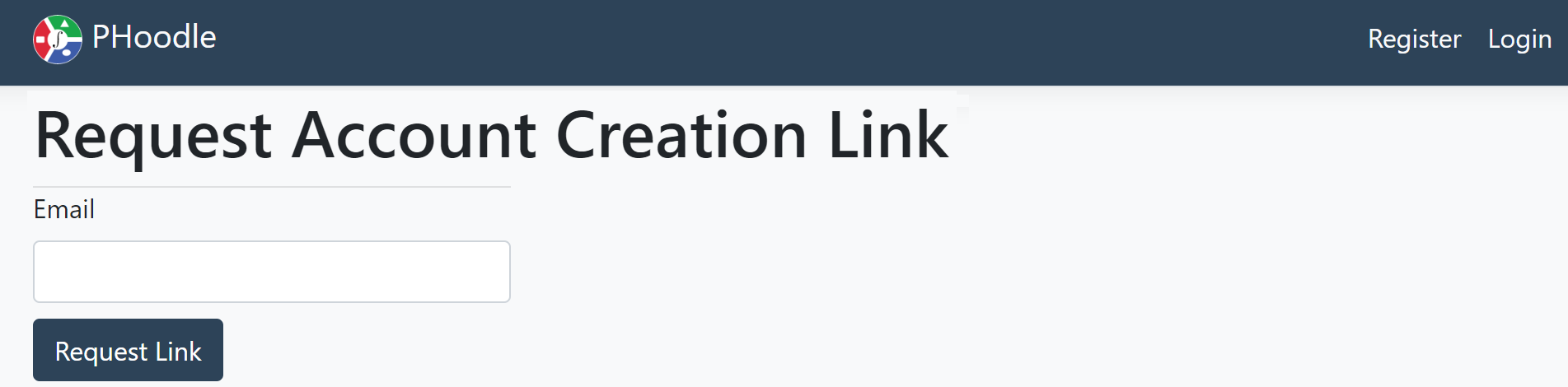}\caption{\label{fig:Registration}Registration}
\end{figure}

In the PHILharmonicFlows engine, this action causes the creation of
an instance of the \emph{Person} user object. Furthermore, the e-mail
address the subject entered is set as an attribute value of the new
object instance. This logic is, however, hidden entirely from the
subjects. From their perspective, the registration is identical to
any other typical information system, as can be seen in Fig. \ref{fig:Registration}.

\paragraph{Scenario 2 (Attending a Lecture)}

After completing Scenario 1, the subjects have a new account in a
data model instance prepopulated with object instances for the scenarios.
In Scenario 2, they are asked to log in to their new account, select
the lecture and ``attend'' the lecture. Again, this seems like a
very simple and familiar task to the subjects, even in this fully
generic user interface. After logging in, as long as no \emph{Lecture}
object instance has been selected, the navigation menu shows an overview
over all top level objects that the current \emph{Person}, i.e., the
subject, is allowed to read.

\begin{figure}[H]
\centering{}\includegraphics[width=0.9\columnwidth]{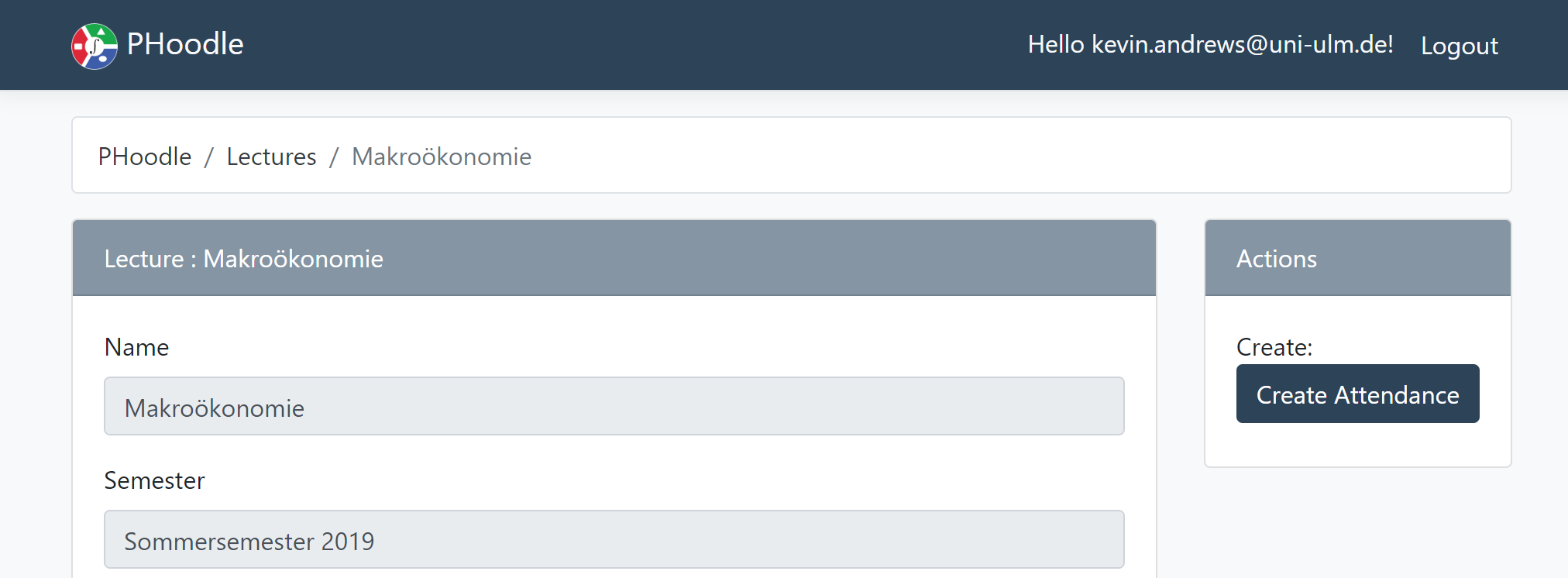}\caption{\label{fig:CreateAttendance}Attending a Lecture}
\end{figure}

Once navigation to a lecture has occurred, the PHILharmonicFlows engine
creates a button for the to-do item \emph{Create Attendance} using
to the concepts presented in Section \ref{subsec:Generating-To-do-Lists}
(cf. Fig. \ref{fig:CreateAttendance}). As the user object instance
that was created in Scenario 1 has no other permissions for the lecture,
except the creation of an attendance, no other actions are offered
by the user interface until this step is completed. To hide the complexity
of the creation of the \emph{Attendance} object instance and as the
relation instances between the \emph{Attendance}, the \emph{Lecture},
and the \emph{Person}, the PHILharmonicFlows engine uses the CORE
algorithm (cf. Algorithm \ref{alg:InstAndLink}) presented in Section
\ref{subsec:Data-Model-Actions}.

\paragraph{Scenario 3 (Checking assigned Tutorial)}

After completing Scenario 2, the subjects are asked to log out and
log back in to the system, to simulate the passing of time between
attending a lecture and having a tutorial assigned by a supervisor.
While the subject is logged out, the system automatically assigns
(hard-coded for the study) the subject to a tutorial by creating the
necessary relations between the \emph{Attendance} object instance
and one of the preexisting \emph{Tutorial} object instances. According
to the principles of object-aware process management, this new relation
path between the \emph{Person} object instance representing the subject
and the \emph{Tutorial} object instance he is assigned to immediately
allows the PHILharmonicFlows process engine to resolve new permissions
for the subject. From the perspective of the subject, the task of
Scenario 3, to check which tutorial he was assigned to, is therefore
simple.

\begin{figure}[H]
\centering{}\includegraphics[width=0.9\columnwidth]{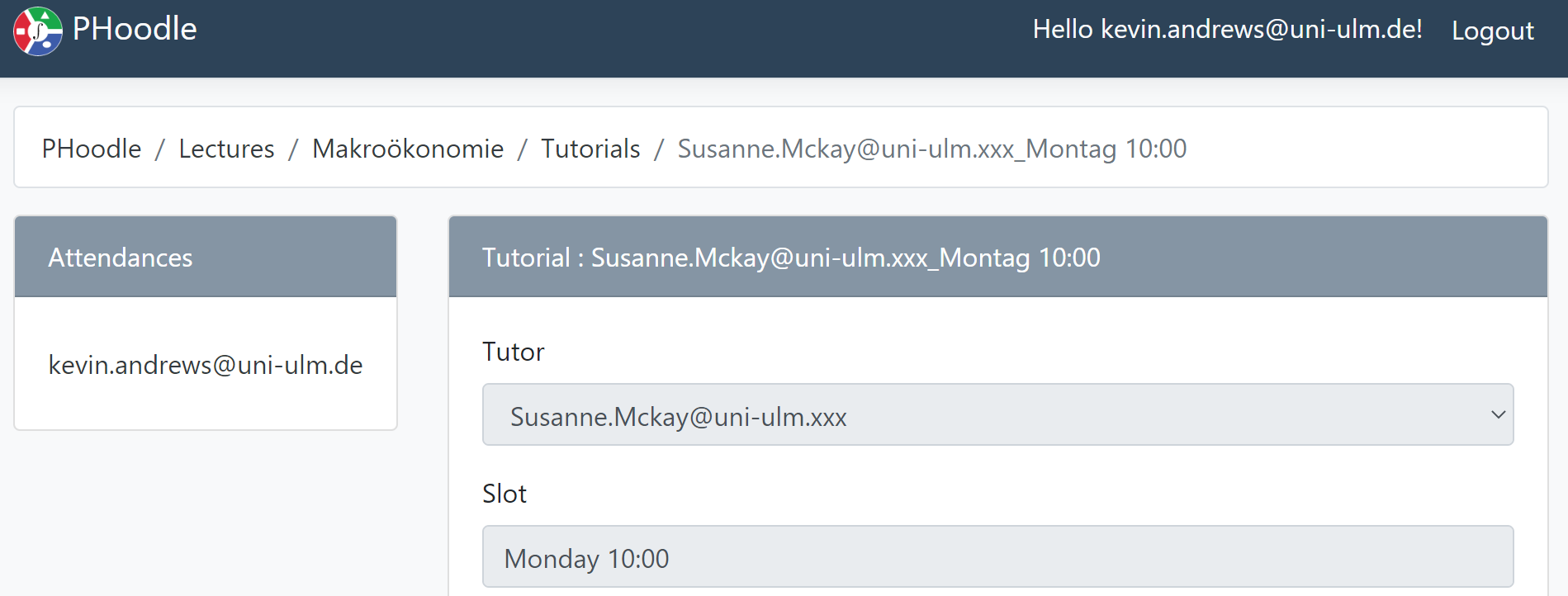}\caption{\label{fig:TutorialDetails}Tutorial Details}
\end{figure}

As the user interface only presents subjects with navigation options
to objects that they have permission to read, they must merely navigate
to the \emph{Lecture} they selected in Scenario 2 and are immediately
presented with a new navigation option to their assigned tutorial.
Clicking the navigation link shows the form for the \emph{Tutorial}
object instance, which includes attribute values that contain relevant
details about the \emph{Tutorial}, such as the \emph{Tutor} and the
concrete time slot in which the tutorial takes place (cf. Fig. \ref{fig:TutorialDetails}).

\paragraph{Scenario 4 (Creating an Exercise Sheet Submission)}

\begin{figure}[H]
\centering{}\includegraphics[width=0.9\columnwidth]{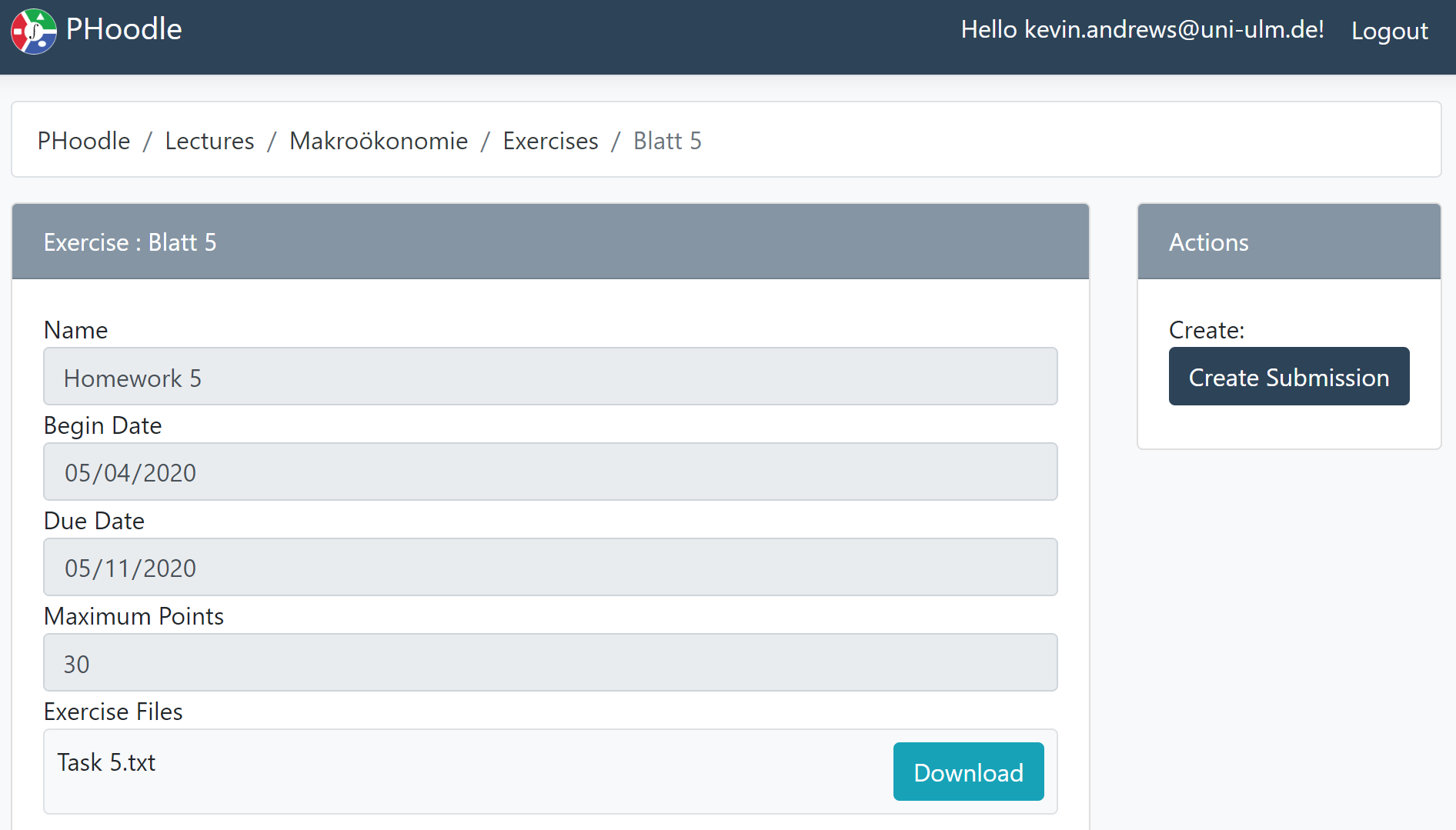}\caption{\label{fig:ExerciseInstance}Exercise Object Instance}
\end{figure}

After the subject completes Scenario 3, the system generates an example
exercise sheet and the subject is asked to download the exercise sheet
and complete the task it describes. The task itself is merely creating
a file with a certain name and uploading it as part of the solution
submission for the exercise sheet. The subjects must first find the
\emph{Exercise} object instance they are supposed to create a \emph{Submission}
for. However, with the help of the To-Do list (or the navigation menu),
this task is fairly simple. Upon selecting the correct object instance,
the form for the Exercise object instance is shown, with the option
to download the PDF file containing the task details (cf. Fig. \ref{fig:ExerciseInstance}).
While drawing the form, the PHILharmonicFlows process engine also
evaluates the permissions of the subject and generates the button
for the \emph{Create Submission} action, which the subject must press
to create a new Submission object instance and (according to the CORE
algorithm) also the relation instances to the correct \emph{Attendance}
and \emph{Exercise} object instances automatically. Finally, while
viewing the form for the new \emph{Submission} object instance, the
lifecycle process for the \emph{Submission} allows the user interface
to generate a form with the correct guiding markings, indicating the
attribute value (in this case the ``solution'' file) the subject
must supply to be able to submit the submission. Supplying the \emph{Solution}
attribute with a value causes an advancement in the lifecycle process
for the object instance that the form is displaying (cf. Fig. \ref{fig:Example-PHILharmonicFlows-Object}).
As soon as the process engine registers the updated attribute value,
the user interface adapts and displays the \emph{Submit} action, allowing
the subject to advance the lifecycle to the next state, thereby completing
the submission.
\begin{figure}[H]
\centering{}\includegraphics[width=0.9\columnwidth]{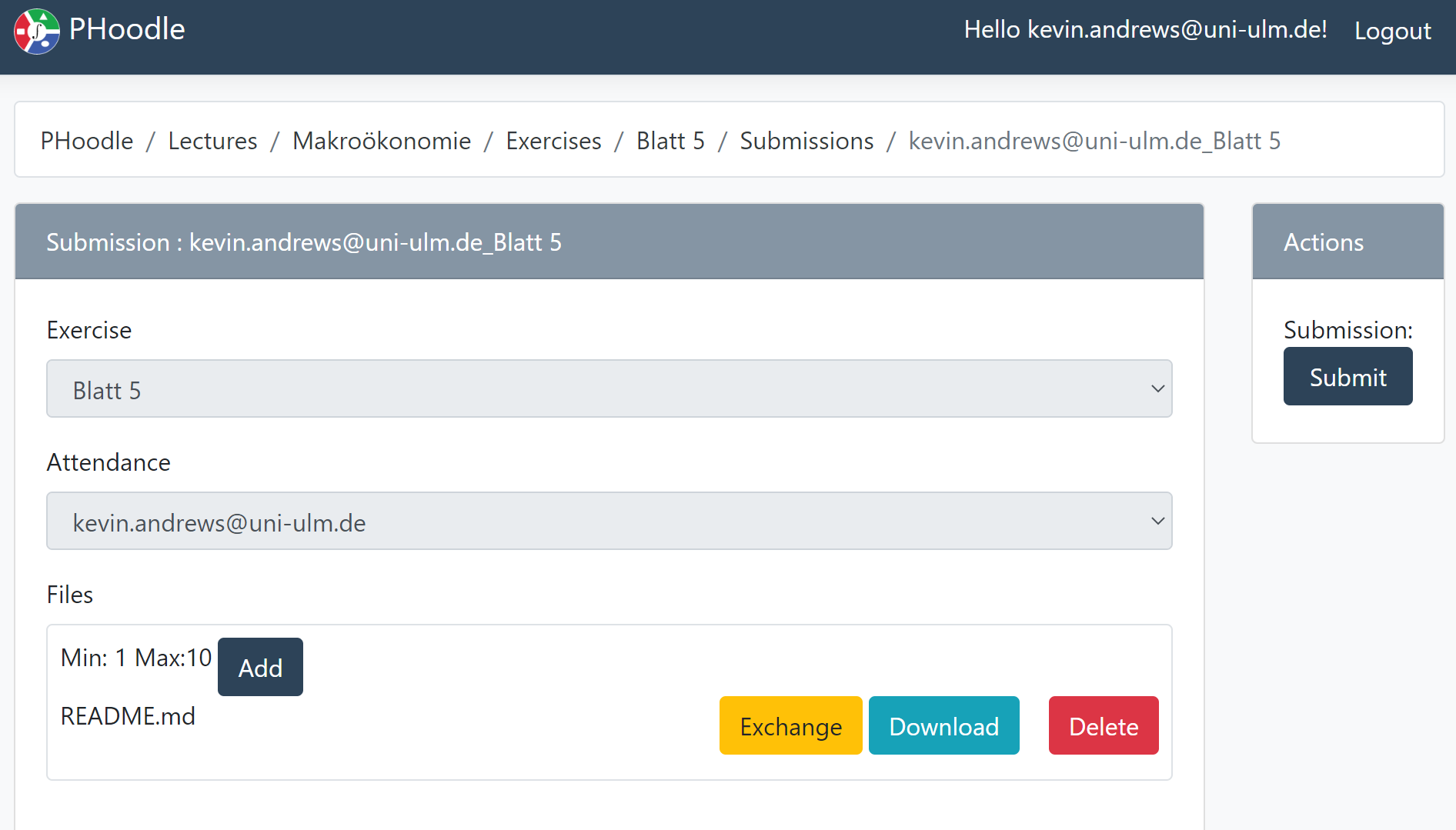}\caption{\label{fig:SubmissionInstance}New Submission Object Instance}
\end{figure}

\paragraph{Scenario 5 (Checking achieved Points)}

\begin{figure}[H]
\centering{}\includegraphics[width=0.9\columnwidth]{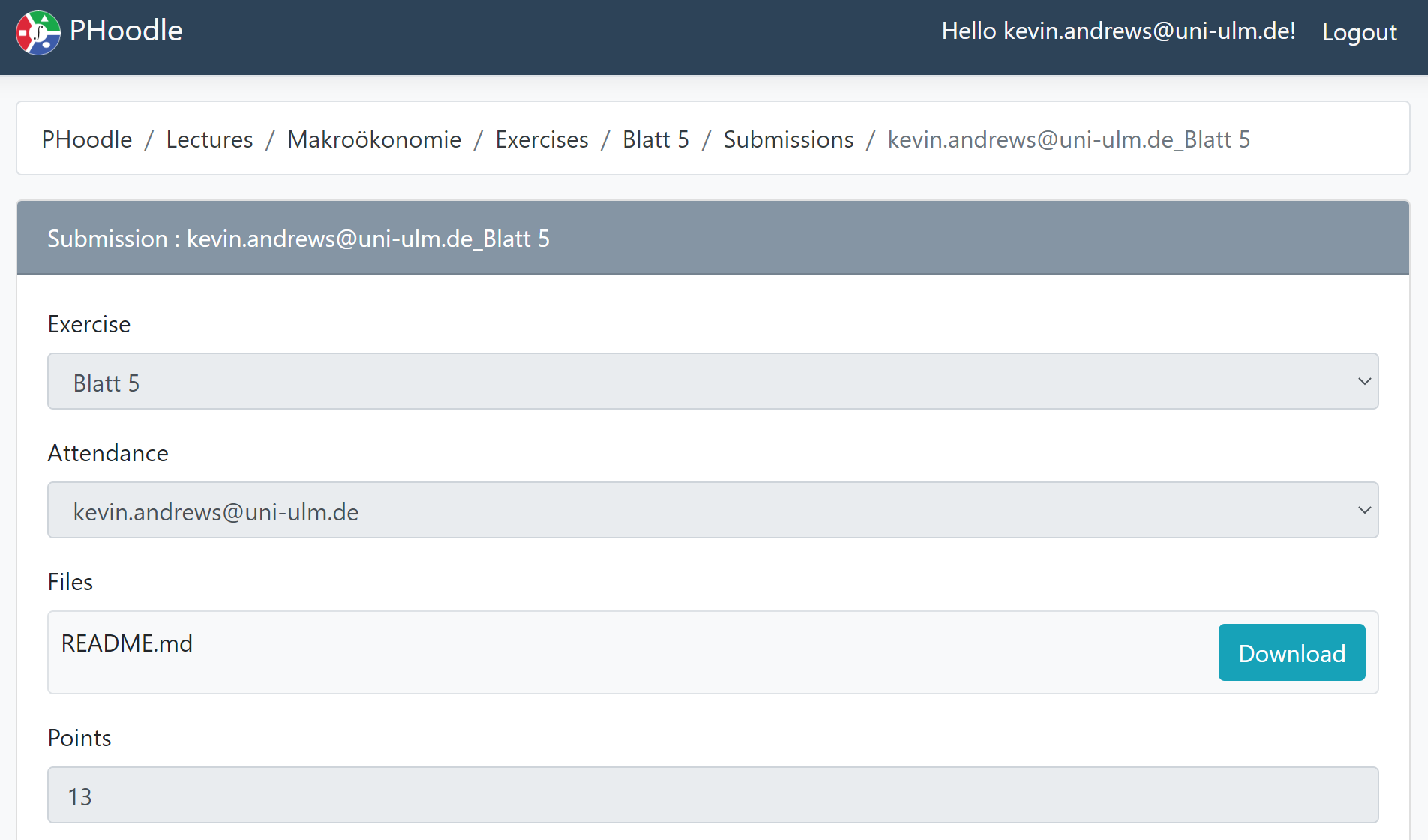}\caption{\label{fig:SubmissionPoints}Submission with Points}
\end{figure}

The final scenario involves asking the subjects to check the number
of points the received as a rating for their submission. In the break
between Scenarios 4 and 5 the system is hard-coded to assign a random
rating to the submission, simulating the actions of a tutor. The difficulty
in this scenario is that \emph{Submissions} are attached either to
\emph{Exercises} or \emph{Attendances}, but not to the \emph{Lecture}
itself. In consequence, in contrast to the other scenarios, where
navigation was merely selecting the \emph{Lecture} and then finding
the object instance in the navigation bar, Scenario 5 involves a multi-level
navigation. Furthermore, as checking the achieved points is not mandatory,
as it is not part of the lifecycle process of a \emph{Submission},
no to-do item is generated for this scenario. Therefore, the subjects
must select either their \emph{Attendance} object instance, which
shows all their submissions, or the correct \emph{Exercise} object
instance, which shows all submissions attached to the exercise, filtered
by their permissions. This corresponds to the examples of object content
navigation given in Section \ref{subsec:Generating-the-Navigation},
i.e., Fig. \ref{fig:filteredoverexercise} and Fig. \ref{fig:filteredoverattendance}.
Once one of the navigation paths is selected by navigating to the
\emph{Attendance} or the \emph{Exercise}, the \emph{Submission} object
instance created in Scenario 4 can be navigated to, allowing the user
interface to display the corresponding form. The form, which still
shows the same lifecycle process as in Scenario 2, is presented with
additional information, including the required \emph{Points} attribute.
This is due to the fact that the subjects have different permissions
in Scenario 5, as the \emph{Submission} object instance was transitioned
to the \emph{Pass }state during the intermission between Scenarios
4 and 5 by the tutor. The resulting form can be seen in Fig. \ref{fig:SubmissionPoints}.

\subsubsection{Study Goals}

The goal of this study was to evaluate the usability of the generic
user-interface of PHILharmonicFlows. In particular, we wanted to gauge
if we had succeeded in abstracting the inherent complexity of object-aware
process management away from the user interface part of the process
management system. Furthermore, as this particular approach to user
interface generation, i.e., based on a data-centric process model,
is novel, we wanted to evaluate whether users could understand and
interact with a completely generic user interface. In this context,
note that the user interface prototype we evaluated does not contain
a single line of code which is specific to the PHoodle process model.
All labels, navigation structure, actions, etc. are derived from the
process model itself.

The only information we provide to the user interface is an internal
data model instance identifier that allows the user interface to request
the correct data model instance on startup. Keeping this in mind,
one of the main goals of the evaluation was to determine whether the
concepts we developed for generating a user interface from the process
model (cf. Section \ref{sec:Presenting-Object-aware-Processe}) were
enough to guide a novice user through the various tasks set forth
by the scenarios. Furthermore, by including the group of experts,
we were able to measure whether a learning curve was necessary to
really master the user interface. The hypothesis was, that if the
novices took significantly longer than the experts to complete the
scenarios or required significantly higher mental effort, the generated
PHoodle user interface must have deficits preventing novice users
from knowing where in the user interface to complete necessary actions.

\subsubsection{After Scenario Questionnaire (ASQ) Results\label{subsec:ASQ-Results}}

\begin{figure}[H]
\subfloat[Mental Effort\label{fig:Mental-Effort}]{\begin{centering}
\includegraphics[width=0.475\columnwidth]{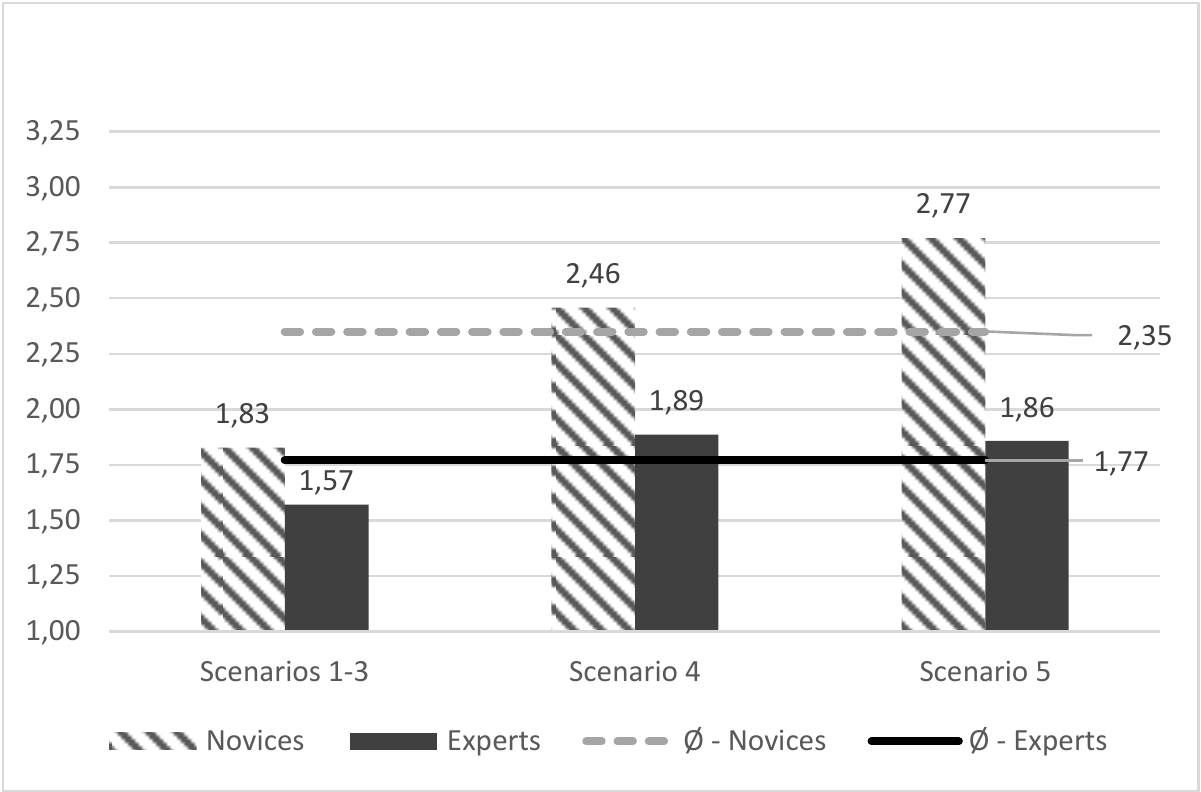}
\par\end{centering}
}\hfill{}\subfloat[Efficiency]{\centering{}\includegraphics[width=0.475\columnwidth]{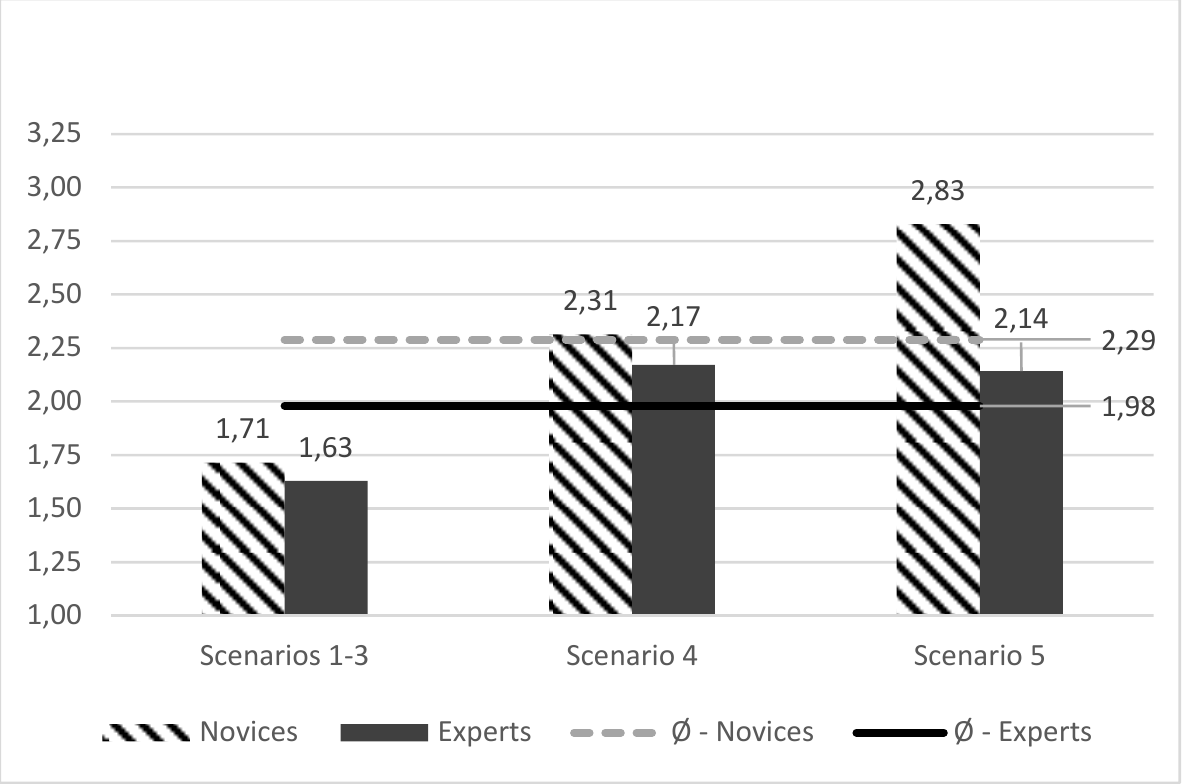}}

\subfloat[Effectiveness]{\centering{}\includegraphics[width=0.475\columnwidth]{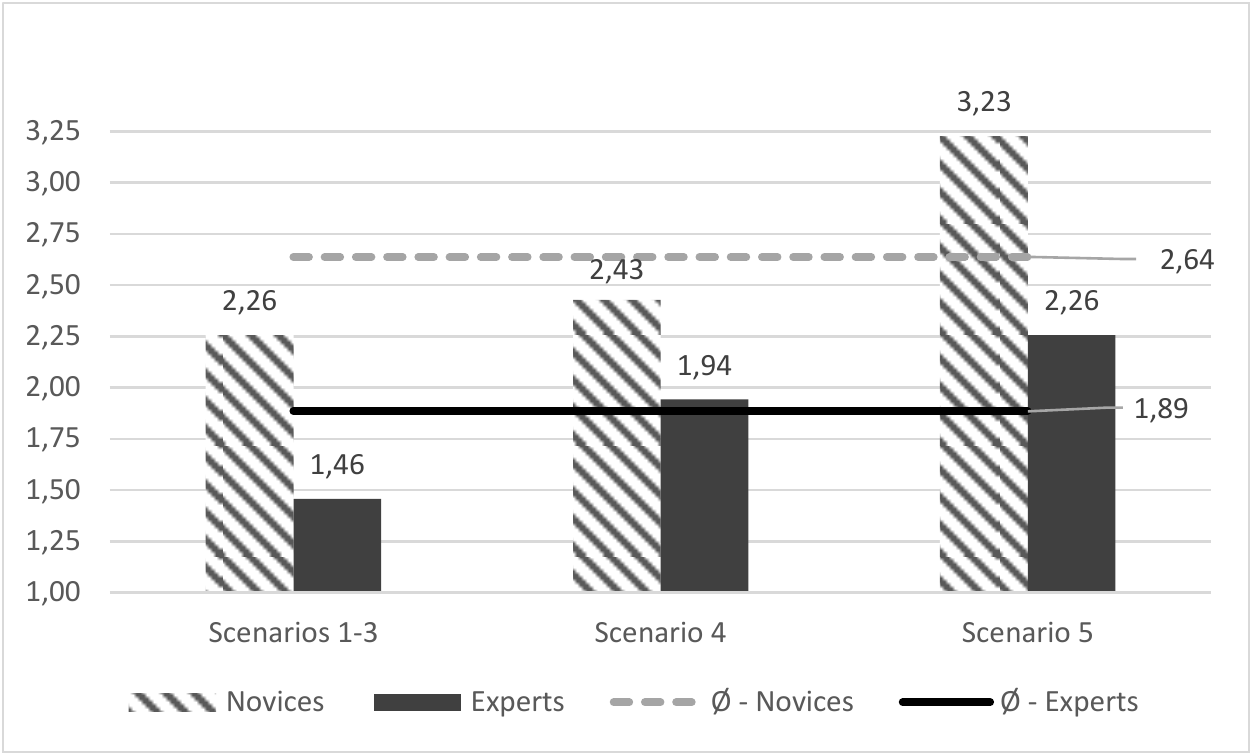}}\caption{ASQ Results\label{fig:ASQ-Results}}
\end{figure}

Upon completion of the more complex Scenarios 3, 4, and 5, subjects
were asked to fill out an After Scenario Questionnaire (ASQ). The
ASQ is an established tool for psychometric evaluation in computer
usability studies \citep{asq}. While the ASQ only has a few carefully
selected questions, they are repeatedly answered by subjects after
each scenario they complete. This immediate evaluation allows for
the fine-grained evaluation of the experience subjects had in a scenario.
The questions posed in our study allow for the evaluation of three
distinct metrics: perceived mental effort, perceived efficiency, and
perceived effectiveness. While the answers to the ASQ are given on
a seven-point Likert scale, the result charts shown in Fig. \ref{fig:ASQ-Results}
are clipped to maximum value of 3.25 to save space. This was possible
as none of the average results of the ASQ was above 3.23. The Likert
scales are structured so that \textbf{lower} values are ``better''
in terms of usability.
The charts in Fig. \ref{fig:ASQ-Results} allow us to extract some
basic observations:
\begin{itemize}
\item The average ASQ results across all 70 subjects and scenarios for the
dimensions Mental Effort $\left(\frac{2.35+1.77}{2}=2.06\right)$,
Efficiency $\left(\frac{2.29+1.98}{2}=2.16\right)$, and Effectiveness
$\left(\frac{2.64+1.89}{2}=2.27\right)$ are all around 2. As the
results could range from values 1 (extremely low/efficient/effective)
to 7 (extremely high/inefficient/ineffective) on the Likert scale,
indicating that subjects, on average, considered themselves to have
required ``very low'' mental effort, to be ``very effective'',
and ``very efficient''. This, in itself, can be considered a success
from usability perspective, considering that users were interacting
with a user interface generated from a process model.
\item In Scenario 5, experts perceived themselves as requiring less mental
effort, being more efficient, and only slightly less effective than
in Scenario 4. The novices, on the other hand, perceived themselves
as requiring significantly more mental effort, and being significantly
less efficient and effective in Scenario 5, compared to Scenario 4.
This does indicate that there is a learning curve in Scenario 5, with
some novice subjects finding it harder to find their submissions due
to the multi-step navigation over the exercise object. However, as
the experts did not rate Scenario 5 any worse than Scenario 4, one
can conclude that the multi-step navigation only poses a issue for
first-time novice users. Clearly, this also depends on the data-model
being structured in way in which navigation makes sense for users.
\item On average, experts perceived themselves as requiring less mental
effort, being more efficient, and more effective across all scenarios
than the novices did.
\item When correlating the previous observation with the system logs, however,
it is interesting to note that the expert subjects actually took longer
on average to complete all five scenarios. As can be seen in Fig.
\ref{fig:Time-Taken-to}, which shows the times the individual subjects
needed to complete all scenarios, the average time that the expert
subjects needed was almost 15 minutes, while the novice subjects needed
an average of nearly 13 minutes. However, considering the individual
times and averages, the difference is not significant from a statistical
perspective. This is an indicator that the generic user interface
is, even while being fully generic, well fitted to its task of guiding
users along their interactions with the process model. In particular,
as shown by these numbers, first-time novice users have no significant
disadvantage in time it takes them to complete the tasks set forth
by the scenarios, as the user interface guides them along the same
paths that expert users take after weeks of use.
\begin{figure}[H]
\subfloat[Individual Novices]{\centering{}\includegraphics[viewport=5bp 50bp 350bp 170bp,clip,width=0.475\columnwidth]{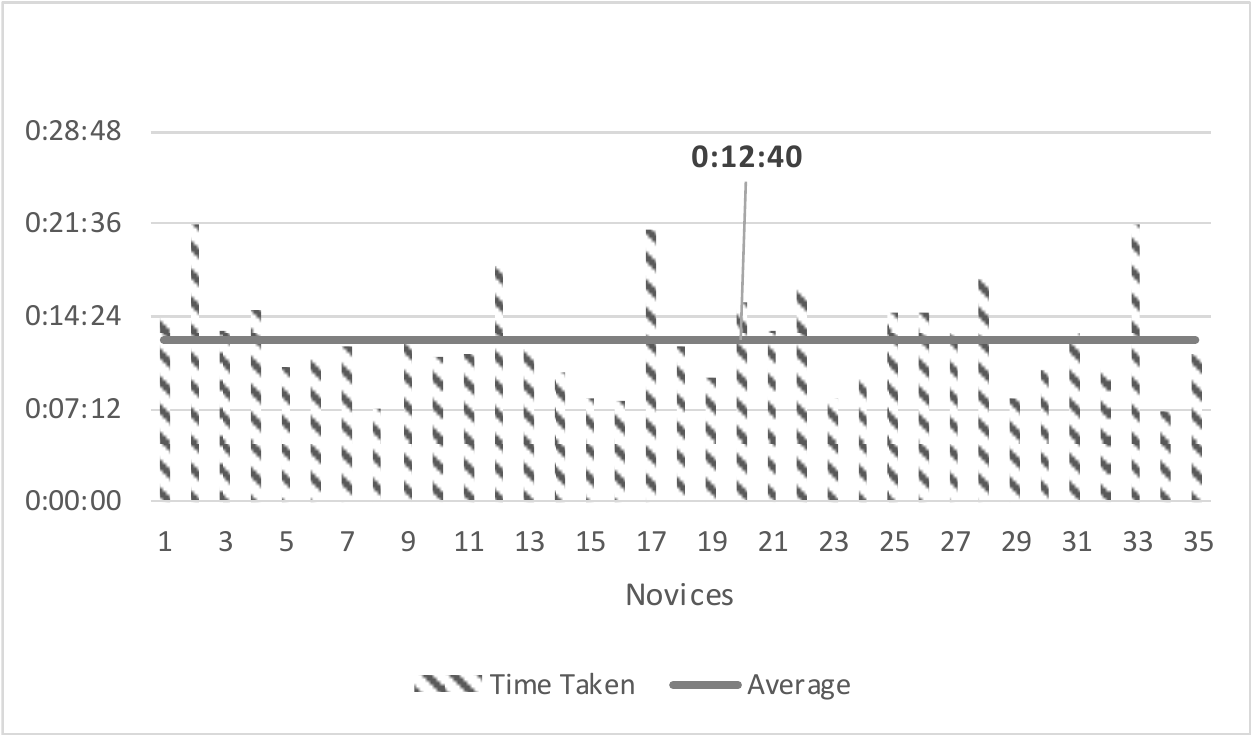}}\hfill{}\subfloat[Individual Experts]{\centering{}\includegraphics[viewport=5bp 50bp 350bp 170bp,clip,width=0.475\columnwidth]{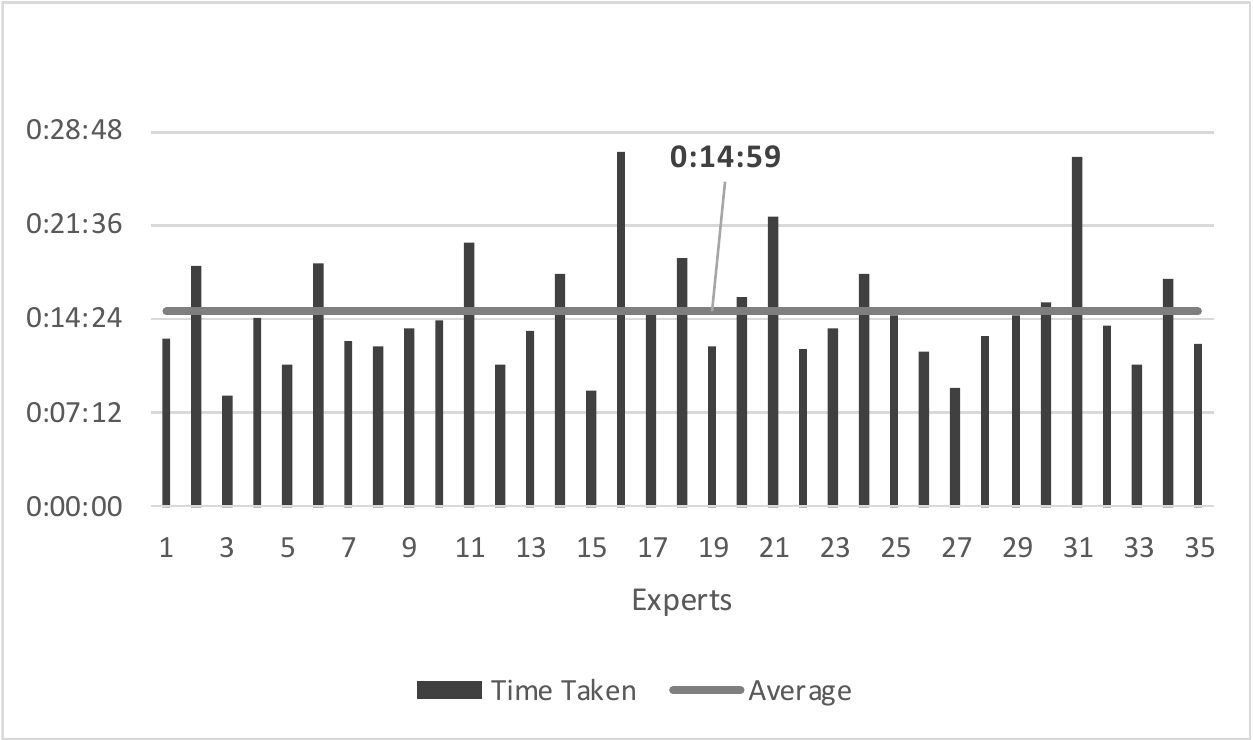}}\caption{Time Taken to Complete all Scenarios\label{fig:Time-Taken-to}}
\end{figure}
\end{itemize}

\subsubsection{ISO-Norm 9241/110 Results\label{subsec:ISO-Results}}

The International Organization for Standardization norm 9241/110 \citep{iso}
concerns the ergonomics of interactive systems and helps identify
usability problems. The norm includes a standardized questionnaire
which was presented to subjects during our evaluation study. In contrast
to the ASQ, the ISO 9241/110 questionnaire was only presented once
to subjects, after completion of all scenarios. In consequence, there
is only one set of results which does not differentiate between the
various scenarios.

\begin{figure}[H]
\centering{}\includegraphics[width=0.7\columnwidth]{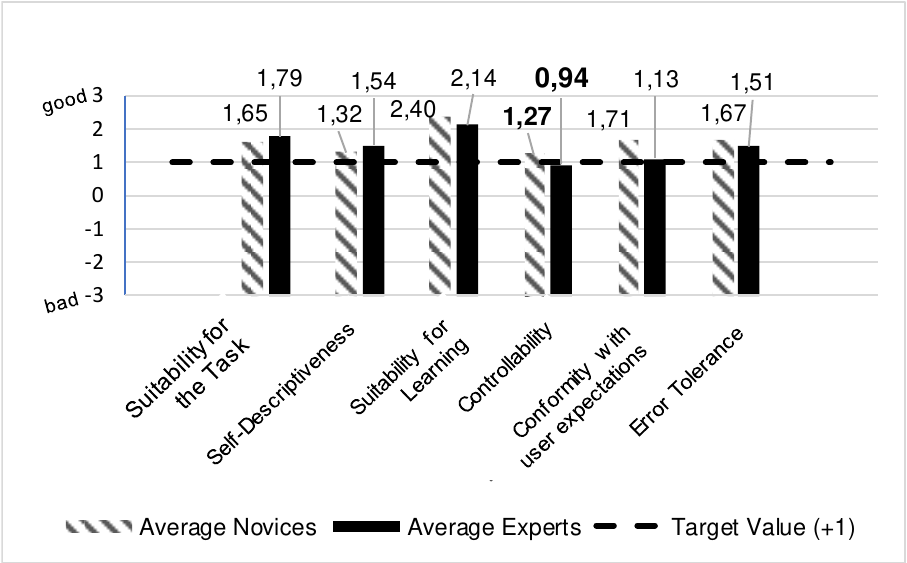}\caption{ISO Norm 9241/110 Questionnaire Results\label{fig:ISO-Norm-9241/110}}
\end{figure}

Of the seven so-called ``aspects'' examined by the ISO 9241/110
questionnaire (\emph{Suitability for the Task}, \emph{Self-Descriptiveness},
\emph{Suitability for Learning}, \emph{Controllability}, \emph{Conformance
with User Expectations}, \emph{Error Tolerance}, and \emph{Suitability
for Individualization}) only six were examined, as \emph{Suitability
for Individualization} was not applicable because the user interface
offers no related options. For each aspect, a series of questions,
16 in total, were answered by subjects. The answers were given on
a seven-point Likert scale, as required by the norm. The evaluation
of an ISO 9241/110 questionnaire is also standardized by the norm,
resulting in the chart shown in Fig. \ref{fig:ISO-Norm-9241/110}.
All results are normalized to the range {[}-3, +3{]}, with the norm
considering aspects that are rated at least +1 to be ``good''. Therefore,
software with at least a +1 rating in all aspects is considered to
have good usability according to ISO 9241/110.
Examining Fig. \ref{fig:ISO-Norm-9241/110} shows that the generic
PHILharmonicFlows user interface hits the target value of +1 across
the board when considering the average ratings across novices and
experts. However, while the average rating for the \emph{Controllability}
aspect is $\left(\frac{1.27+0.94}{2}=1.11\right)$, the experts rated
the aspect below the target value, at 0.94 (cf. bold values in Fig.
\ref{fig:ISO-Norm-9241/110}). The \emph{Controllability} aspect,
which rates whether users thought that they ``felt in control while
working with the software'', is, obviously, a problematic one for
any process management system. While novice users might be thankful
for the guidance they receive when following the execution path predefined
in the process model, experts may feel that their choices are limited
by these boundaries. The experts rating this aspect just shy of ``good''
in the study shows that the problem of perceived lack of control for
expert users does also exist in object-aware processes. However, it
is not nearly as pronounced as it is in activity-centric process management,
where the exact order of all activities is predefined.

The most important aspects for this evaluation are the ratings for
the aspects \emph{Suitability for the Task} and \emph{Self-Descriptiveness}.
They rate whether users thought that ``the software supported them
in completing their tasks'' and ``the software was easily understandable
and self-explanatory'', respectively. According to Fig. \ref{fig:ISO-Norm-9241/110},
both aspects were rated well above the target value of +1, by novices
and experts alike. We consider a good rating in these aspects paramount
to the user experience and, therefore, also acceptance of a generic
user interface created using a certain approach, such as object-aware
process management. Having achieved these results with a prototype
shows the viability of the object-aware process management approach
for generating user interfaces.

\subsubsection{User Experience Questionnaire (UEQ) Results\label{subsec:UEQ-Results}}

\begin{figure}[H]
\centering{}\includegraphics[width=1\columnwidth]{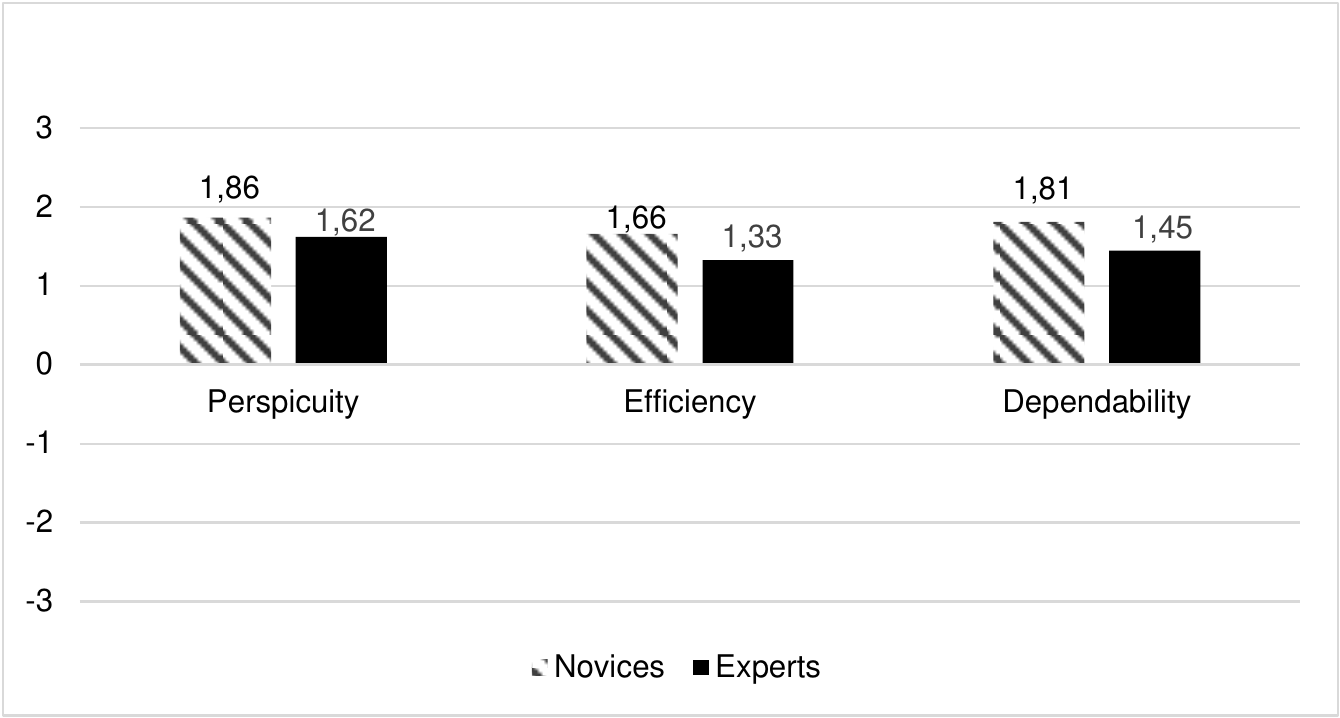}\caption{UEQ Results\label{fig:UEQ-Results}}
\end{figure}

\begin{figure}[H]
\centering{}\includegraphics[width=1\columnwidth]{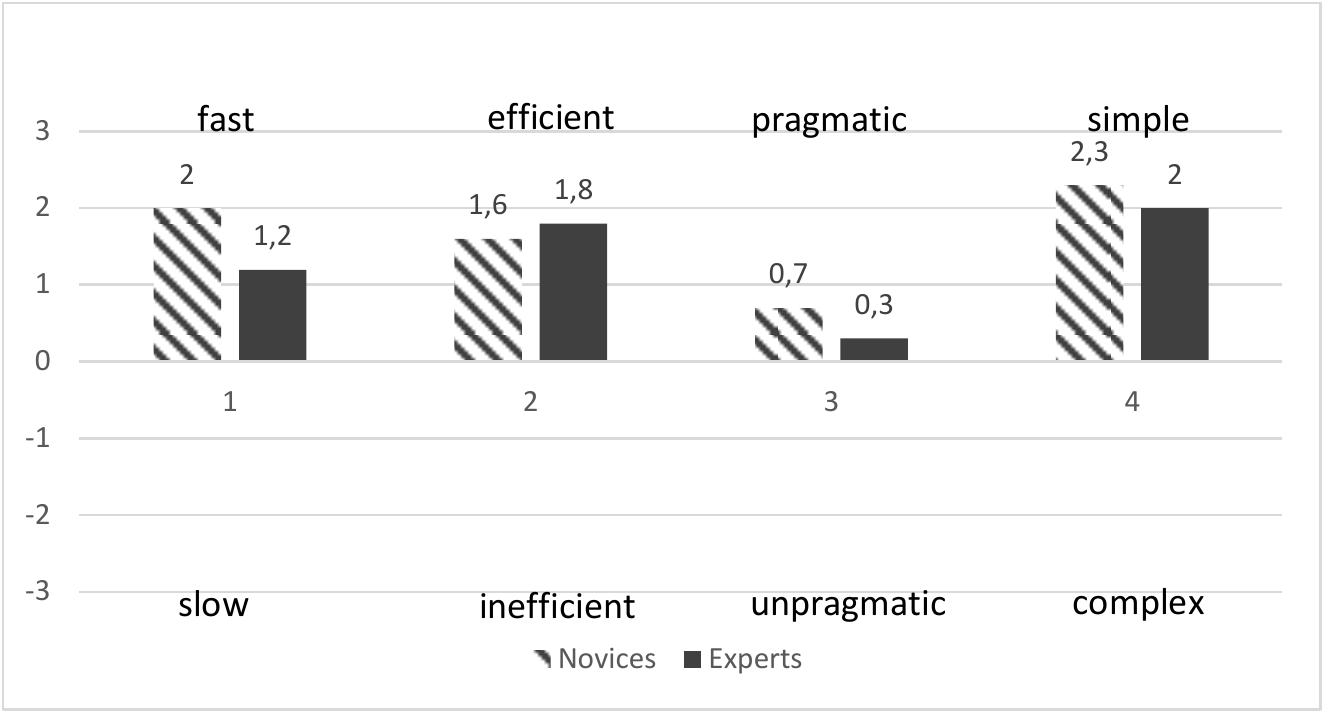}\caption{UEQ Dimension: Efficiency\label{fig:UEQ-Dimension:-Efficiency}}
\end{figure}

The final questionnaire presented to subjects was the User Experience
Questionnaire (UEQ) \citep{ueq}. The UEQ is similar to the ISO 9241/110
questionnaire, with seven-point Likert ratings for questions across
six so-called ``dimensions''. However, the UEQ aims at gathering
a quicker, simpler, and more straightforward notion of the experience
users have with a piece of software. The study evaluated three of
the six UEQ dimensions, i.e., \emph{Perspicuity}, \emph{Efficiency},
and \emph{Dependability}, as the other three, i.e., \emph{Attractiveness},
\emph{Stimulation}, and \emph{Novelty}, where not were not goals of
the work presented in this article. Similarly to the ISO 9241/110,
the UEQ suggests a target value for each dimension, which is fixed
at 0.8, meaning that values over 0.8 constitute a ``good'' user
experience. The results for the examined dimensions are shown in Fig.
\ref{fig:UEQ-Results}.
Clearly, the results across all dimensions are well above 0.8, with
\emph{Perspicuity} being rated the highest across novices and experts
alike. As \emph{Perspicuity} basically describes how ``easy to grasp''
something is, it is no surprise to see this dimension rated highly,
as the corresponding aspect in the ISO 9241/110 results, i.e. \emph{Suitability
for Learning} was also the highest rated (cf. Fig. \ref{fig:ISO-Norm-9241/110}).
The UEQ however, does also uncover one of the shortcoming of generic
user interfaces, and our approach in particular. It becomes obvious
when examining the UEQ dimension \emph{Efficiency} closer, that the
lowest rated individual item is how ``pragmatic'' the user interface
is (cf. Fig. \ref{fig:UEQ-Dimension:-Efficiency}). 
We examined this result item closer, as it was the only one below
the threshold of 0.8 for both experts and novices. When correlating
the UEQ results with those of the ASQ (cf. Section \ref{subsec:ASQ-Results})
and the textual comments made by subjects it becomes clear that the
multi-level navigation between related objects, which was necessary
to complete Scenario 5, was perceived as less pragmatic than other
elements of the user interface. While it is a simple concept from
a user perspective and subjects dealt with it efficiently, many subjects
wanted to know why the navigation to individual submissions had not
been placed on the same level as exercises and tutorials. While it
is necessary from a data model perspective to have \emph{Submission}
as a lower level object attached to \emph{Exercise} for various conceptual
reasons (cf. Section \ref{sec:Fundamentals}), this is, naturally,
completely unclear to users who just want to know why they cant navigate
directly to their submissions when viewing the details of a lecture
in their e-learning system.

This is a limitation of our approach to generic user interfaces, as
we, by design, do not offer any configuration options for the user
interface that are not part of the process model, as this would erode
a core aspect of our approach. In particular, adding a bunch of necessary
UI configuration on top of the process model would defeat the purpose
of having a fully generically generated user interface, as it would
then no longer be fully generic.

\subsection{Real-World Deployment Measurements\label{subsec:Real-World-Deployment-Measuremen}}

\begin{figure}[H]
\centering{}\includegraphics[width=1\columnwidth]{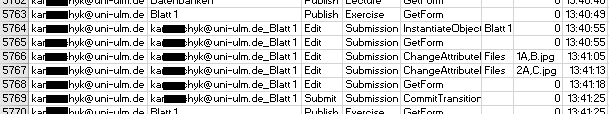}\caption{\label{fig:logs}Sample Log Entries (redacted to comply with GDPR)}
\end{figure}

This evaluation is based on a full-scale deployment of the PHoodle
data model in the PHILharmonicFlows engine at Ulm University in the
summer semester of 2019. The PHoodle deployment\footnote{Feel free to log in to the live instance at \url{https://phoodle.dbis.info}

Username: \emph{edoc.demo@uni-ulm.}de Password: \emph{edoc.demo}} replaced the established Moodle e-learning platform over an entire
semester (exactly 100 days) for hundreds of students. We were able
to parse and extract information from nearly 70,000 PHILharmonicFlows
process engine log entries which were cleaned and aggregated to a
log containing 39,890 interactions. To be precise, each entry in this
interaction log is exactly one click or action by a real student,
tutor, or supervisor. The log entries we removed from the initial
\textasciitilde 70,000 were system debugging information and related
log entries we had to create to ensure we could react to problems
in the live system if any arose. A sample of some of the columns in
the the raw log output we gathered for analysis is shown in Fig. \ref{fig:logs}.

Note that there are a large number of columns missing, as all labels
shown in Fig. \ref{fig:logs}, such as the names of users, objects,
attributes, etc. are actually logged as 64bit integer reference values
internally. These reference id columns are then supplemented by a
``pretty'' string-based form for manual log examination in Excel,
which is what is displayed in Fig. \ref{fig:logs}. Furthermore, all
user names and file names etc. were redacted to comply with the GDPR.

We are able to further aggregate this data using common techniques
such as pivot tables and lookup formulas, providing us with some useful
metrics which show the extent to which the process was utilized in
our real-world deployment. We provide some of these to give the reader
a sense of scale of our deployment in Table \ref{tab:PHoodle-Deployment-Metrics}.
\begin{table}[H]
\centering{}%
\begin{tabular*}{1\columnwidth}{@{\extracolsep{\fill}}|c|>{\centering}b{0.125\columnwidth}|>{\centering}b{0.15\columnwidth}|>{\centering}b{0.175\columnwidth}|>{\centering}b{0.15\columnwidth}|}
\hline 
Object & Amount\\
Instances & Amount\\
 Interactions & Average\\
 Interactions\\
per Instance & Max\\
 Interactions\\
per Day\tabularnewline
\hline 
\hline 
Attendance & 137 & 3233 & 23.6 & 583\tabularnewline
\hline 
Download & 14 & 4574 & 326.7 & 167\tabularnewline
\hline 
Employee & 2 & 14 & 7.0 & 8\tabularnewline
\hline 
Exercise & 5 & 7323 & 1464.6 & 384\tabularnewline
\hline 
Lecture & 1 & 11741 & 11741.0 & 428\tabularnewline
\hline 
Person & 133 & 290 & 2.2 & 136\tabularnewline
\hline 
Submission & 498 & 10689 & 21.5 & 859\tabularnewline
\hline 
Tutor & 6 & 116 & 19.3 & 27\tabularnewline
\hline 
Tutorial & 52 & 1910 & 36.7 & 527\tabularnewline
\hline 
\hline 
All Objects & 848 & 39890 & 47.0 & 1827\tabularnewline
\hline 
\end{tabular*}\caption{PHoodle Deployment Metrics over first 100 Days\label{tab:PHoodle-Deployment-Metrics}}
\end{table}
The logs provide an opportunity to gather any number of other insights
as well, such as the time various object instances spent in certain
states. This can be calculated by examining the first log entry in
which an object instance was in a certain state and subtracting the
timestamp from the first log entry in which the object instance was
in a subsequent state. This allows for interesting insights into the
execution of the process model without any additional configuration.
An example of this is given by the detailed overview over the time
metrics for the \emph{Submission} object, as shown in Table \ref{tab:Time-Metrics-for}.
In particular, Table \ref{tab:Time-Metrics-for} shows the minimum,
maximum, average, and median time the 498 submission object instances
spent in the various possible states. Note that the times are not
recorded for state \emph{Rated}\footnote{A combination of the \emph{Pass} and \emph{Fail} states seen in the
slightly simplified lifecycle process example from Fig. \ref{fig:Example-PHILharmonicFlows-Object}.}, as object instances never leave their final state, i.e., all instances
were still in state \emph{Rated} when the log was extracted.
\begin{table}[H]
\centering{}%
\begin{tabular}{|c|c|c|c|c|}
\hline 
State & Min Time & Max Time & Average Time & Median Time\tabularnewline
\hline 
\hline 
Edit & 00:00:02 & 498:29:12 & 08:12:51 & 00:00:53\tabularnewline
\hline 
Submit & 00:00:51 & 535:23:54 & 130:41:14 & 106:16:21\tabularnewline
\hline 
Rate & 0:00:18 & 668:10:29 & 102:43:05 & 66:49:51\tabularnewline
\hline 
Rated & - & - & - & -\tabularnewline
\hline 
\end{tabular}\caption{Time Metrics for \emph{Submission}\label{tab:Time-Metrics-for}}
\end{table}
Finally, it is noteworthy that such examinations can be completed
in a generic fashion as well. No model-specific configuration necessary
to enable these kind of analyses, which are possible at any point
in time during deployment and execution of a PHILharmonicFlows data
model. This is possible due to the fact that the process engine produces
log entries for each external interaction, regardless of process model.
These detailed logs are another advantage of the fine-grained modeling
approach that object-aware processes dictate. The modeling of forms
and navigation, by means of the lifecycle processes and data model
at design-time, ensures that all navigation and form related operations
are transparent and correlatable by the process engine at runtime.
In turn, this enables them to be logged in a generic fashion for later
analysis. Even the fairly simple Table \ref{tab:Time-Metrics-for}
offers a great deal of information, such as that the median time students
took to edit their submissions was 53 seconds. On the other hand,
one can observe that the average time is skewed massively by the fact
that a small handful of students had created submissions and then
not submitted them for many days, which is perfectly legitimate due
to the flexible nature or object-aware processes.

Such analyses can go even more into extreme detail, such as that the
average time it took lecture supervisors between entering the name
and the maximum points for an exercise sheet was 79 seconds. Finally,
any number of analyses can also be completed using the actual data
values present in the attributes of the object instances, ranging
from the very useful, such as the average points submissions achieved
for each exercise sheet, to the not so useful, such as the average
amount of points tutors awarded to submissions, grouped by the day
of the week on which the rating was given.

\begin{figure}[H]
\begin{centering}
\includegraphics[width=1\columnwidth]{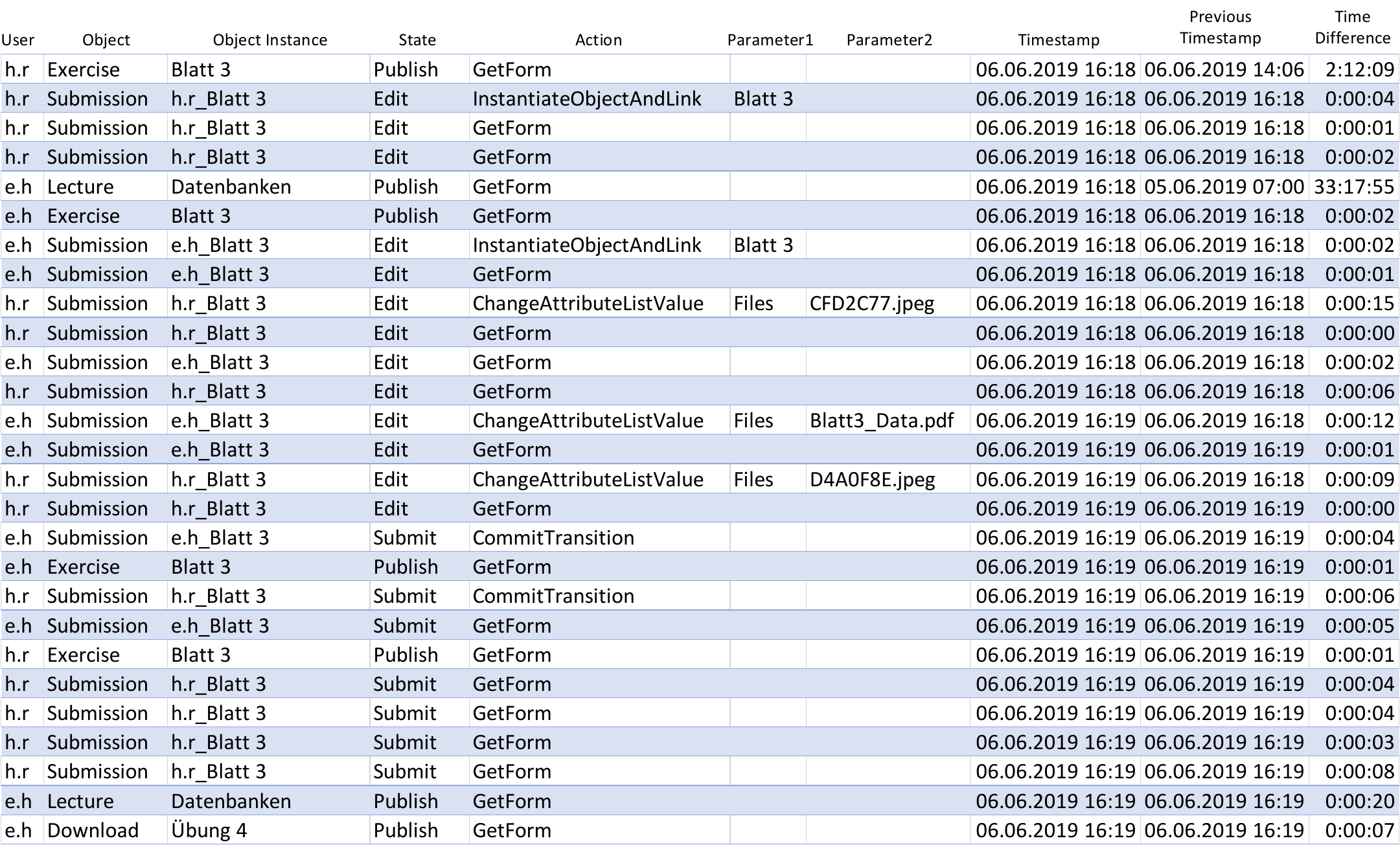}
\par\end{centering}
\caption{\label{fig:logs-2}Sample Navigation and Interaction Chain (redacted)}
\end{figure}

One of the main things we examined for this evaluation was the navigation
and interaction chains that could be identified in the logs, as we
wanted to ensure that users were not aimlessly searching for information
while interacting with the user interface. To this end we wrote a
series of macros that analyzed individual chains such as those shown
in Fig. \ref{fig:logs-2}. The additional columns \emph{Previous Timestamp}
and \emph{Time Difference} were added to the raw log via formulas.
Essentially, the \emph{Previous Timestamp} column points to the previous
timestamp what was caused by the same user, while \emph{Time Difference}
calculates the difference between the current and previous timestamp
for said user. As the \emph{Time Difference} column allows for the
identification of large gaps between user interactions, it can be
utilized for grouping log entries into coherent user interactions
chains.

Assuming a threshold of 2-3 minutes between two log entries as an
indicator that a user started a new interaction chain, the excerpt
from the log shown in Fig \ref{fig:logs-2} shows two clear interaction
chains from two different users, \emph{h.r} and \emph{e.h} (redacted
to initials). Note that these interaction chains occurred concurrently.
Clearly, the starting points of the interaction chains are the row
in which \emph{h.r} has a time difference between timestamps of 2:12:09
(i.e., over 2 hours of inaction), and the row in which \emph{e.h}
has a time difference of 33:17:55 (i.e., over 33 hours of inaction).
The interaction chain of \emph{h.r} starts with a navigation to object
instance ``Blatt 3\footnote{German: Exercise Sheet 3}'' indicating
that he or she had bookmarked the exercise sheet in the browser and
navigated to it directly, causing the \emph{GetForm} action to be
logged. Four seconds after loading the form (including page loading
time), \emph{h.r} clicked on the ``Create Submission'' button generated
by the user interface when viewing an exercise sheet. This caused
the user interface to execute the \emph{InstantiateObjectAndLink }method
in the process engine, which calls the CORE algorithm (cf. Alg. \ref{alg:InstAndLink})
and creates a new \emph{Submission} object instance and the necessary
relations to the \emph{Exercise }object instance ``Blatt 3'' and
the \emph{Attendance} object instance belonging to the user ``h.r''.
Finally, \emph{GetForm }is executed for the new \emph{Submission}
to display it in the browser. Note that \emph{GetForm} is logged twice
when navigating to new object instances due to a technical limitation.

At this point, the second interaction chain, belonging to user \emph{e.h},
starts concurrently. It takes \emph{e.h} a mere two seconds (including
page loading time) after navigating to the lecture ``Datenbanken\footnote{German: Databases}''
to navigate to the correct exercise and another two seconds to create
the submission using the generated action button. Only 12 seconds
later the engine logged \emph{e.h} adding the file ``Blatt3\_Data.pdf''
to the \emph{Files} attribute of the new \emph{Submission} object
instance. Four seconds later \emph{e.h} clicked the ``Submit'' button
that was generated by the user interface as a response to all steps
in the lifecycle process of state \emph{Edit} being provided with
values. Note that this action is logged under the internal name ``CommitTransition'',
which marks the transition of an object instance to a new state, in
this case ``Submit''. Following the \emph{e.h} chain further shows
that \emph{e.h} proceeded to navigate back to the \emph{Exercise},
then forwards again to the submission, presumably to check if it had
been submitted correctly (students were told that PHoodle was a prototype
after all). Finally, the chain ends after two more log entries showing
that \emph{e.h} navigated \emph{up} from his or her \emph{Submission}
to the ``Datenbanken'' \emph{Lecture} (skipping the ``Blatt3''
\emph{Exercise}, i.e. presumably via direct breadcrumb navigation,
cf. Fig. \ref{fig:SubmissionPoints}) and then \emph{down} into a
\emph{Download} object instance. All the navigation paths made visible
by these logs, both up to higher levels and down to lower levels,
can be traced back to the data model shown in \ref{fig:relationpaths}.
Furthermore, this allows us to create statistics based on how long
users needed to perform individual actions and entire interaction
chains. A few of these statistics, such as that it took students an
average of just below 23 seconds to upload a submission, are helpful
in allowing us to gauge whether the PHILharmonicFlows user interface
presented in this article was as simple and efficient to use as the
study we conducted in Section \ref{subsec:Empirical-User-Experience}
suggests. In particular, we examined the data and extracted the exact
time it took students between having created a \emph{Submission} object
instance and providing values for the \emph{Files} attribute instance
and grouped the times by the \emph{Exercises} object instances the
\emph{Submission} object instances are related to. The results were
that students had completed this step in an average of 28 seconds
for exercise sheet 1, 22 seconds for exercise sheet 2, 20 seconds
for exercise sheet 3, 19 seconds for exercise sheet 4, and 22 seconds
for exercise sheet 5. This clearly shows that a) the students had
a clear learning effect after having completed the task for the first
time and b) the user interface allowed them to complete their submissions
very quickly, regardless of their level of expertise. This is supported
by the fact that across all 1163 file uploads the measured learning
effect is only statistically significant between the files associated
with submissions to the first exercise sheet (\textasciitilde 28
seconds) when compared to all other exercise sheets (\textasciitilde 19-22
seconds).

In essence, the correlation possibilities and questions that may be
answered by examining the PHILharmonicFlows engine logs are manifold,
and, together with the empirical study, provide us with the certainty
that the users in our real-world deployment could complete their tasks
in a timely manner and without having to re-do steps due to process-related
limitations. Furthermore, the logs we examined, and the generic logging
system in general, provide ample opportunity for future research into
topics such as process mining and conformance checking of data-centric
processes.

\subsection{Discussion and Further Evaluation}

So far, we have made one change to the PHoodle process model based
on the collected data, which we use to identify limitations of the
current concept. In particular, we added a relation between the \emph{Tutorial}
object and the \emph{Lecture} object, which is already reflected in
Fig \ref{fig:Design-time-data-model}. This is due to the fact that
our concept, as explained in Section \ref{subsec:Generating-the-Navigation},
only displays the directly related object instances for a specific
context object. As there is no structural reason to have a relation
between a \emph{Tutorial} and a \emph{Lecture}, as they are already
transitively related via \emph{Tutor}, the relation was left out of
the initial model. However, users expect to be able to see their assigned
\emph{Tutorial} when viewing a \emph{Lecture}, which led to confusion
as the users had to first select their \emph{Tutor} to see their assigned
\emph{Tutorial}. Requiring these minor changes to an object-aware
process model, in addition to the necessity of adding relation attributes
to enable automatic relation creation, are threats to validity of
the solution, as this imposes some constraints on the model for the
user interface to function optimally from a usability perspective.
Nevertheless, even without \textbf{any} changes, the user interface
can be used to interact with any object-aware process, albeit lacking
some ``comfort'' features.

\begin{figure}[H]
\begin{centering}
\includegraphics[width=0.4\columnwidth]{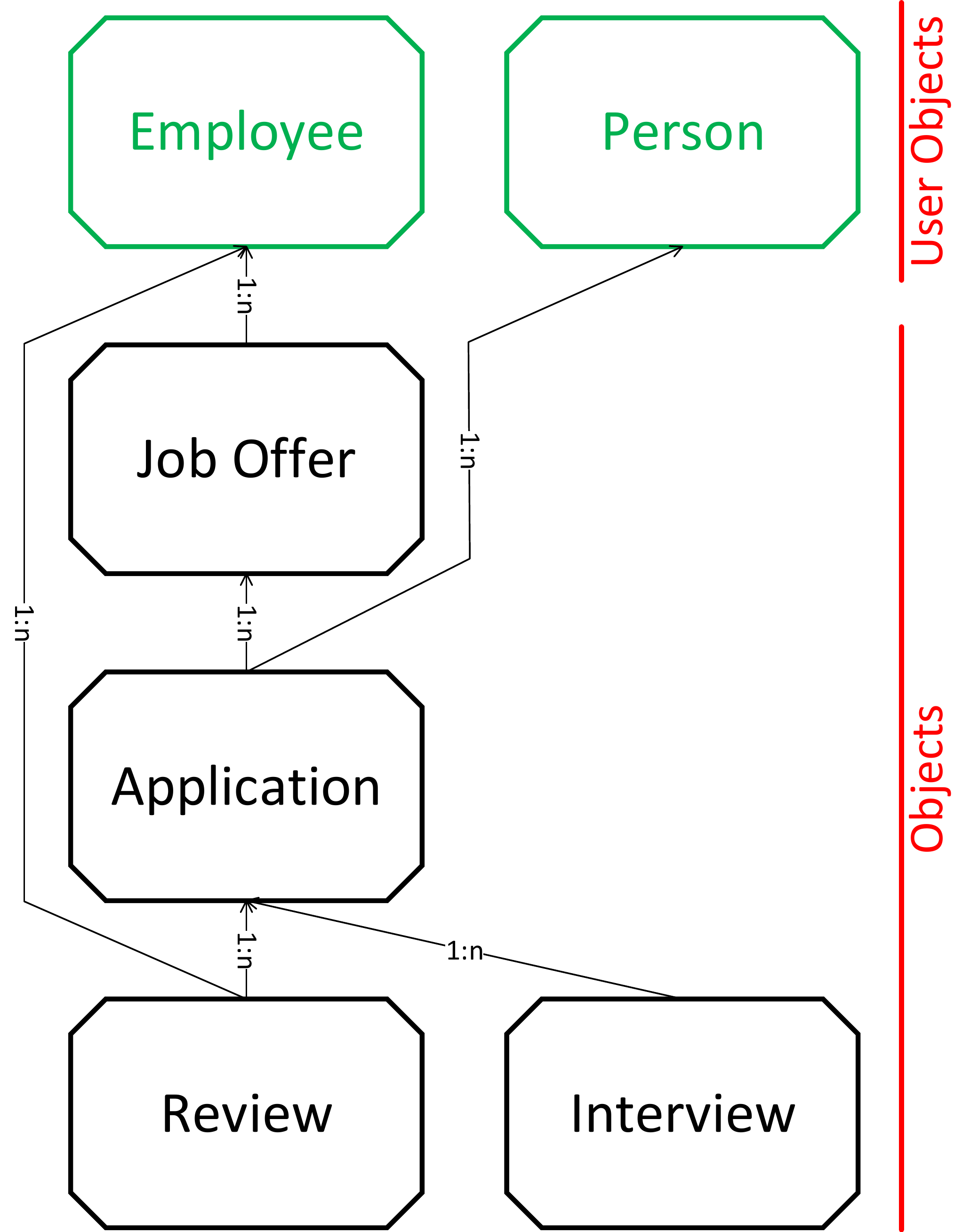}
\par\end{centering}
\caption{\label{fig:Recruitment-Data-Model}``Recruitment'' Data Model}
\end{figure}

\begin{figure}[H]
\begin{centering}
\includegraphics[width=0.7\columnwidth]{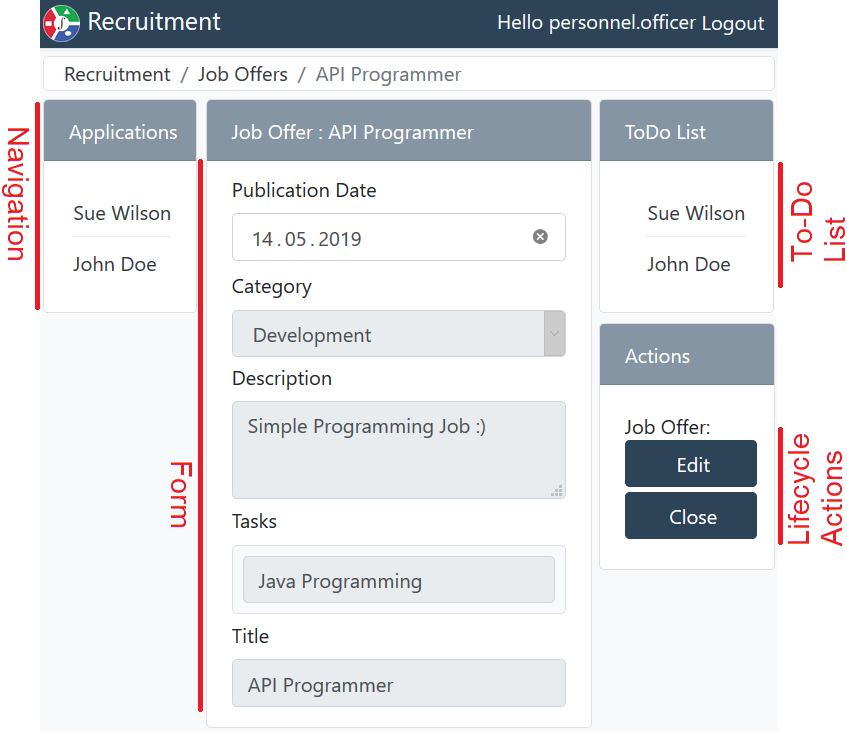}
\par\end{centering}
\caption{\label{fig:recruitmentoverview}View on a Job Offer with Pending Applications}
\end{figure}

To verify that the presented concepts not only work in the context
of the PHoodle e-learning process model, we created a number of different
models employing the concepts. One of these is a human resource management
system that specifically deals with the recruitment of new job applicants.
Without going into too much detail, the data model of the recruitment
example consists of the objects \emph{Job Offer}, \emph{Application},
\emph{Interview}, \emph{Review}, \emph{Person}, and \emph{Employee},
which are related to each other as shown in Fig. \ref{fig:Recruitment-Data-Model}.
Once loaded into the runtime user interface, we were able to generate
user interfaces such as the one shown in Fig. \ref{fig:recruitmentoverview}
with no additional changes to our concepts or code.

 As can be seen in Fig. \ref{fig:recruitmentoverview}, anyone who
was worked with any other PHILharmonicFlows generated user interface,
such as PHoodle, can immediately grasp what is shown on screen. The
object instance shown is a \emph{Job Offer}, there are currently two
\emph{Applications} directly related to said \emph{Job Offer} instance,
and the actions show possible state changes to the \emph{Close} and
\emph{Edit} states that are part of the \emph{Job Offer} lifecycle
process. Furthermore, this information should also be obvious to persons
who have not yet worked with a PHILharmonicFlows process (like PHoodle)
and have no knowledge whatsoever of any internal details such as the
concept, data models and lifecycle processes. We are currently preparing
a similarly thorough usability evaluation study of the recruitment
model to prove the point that the user interface is self-explanatory
for various models.

\section{Related Work\label{sec:Related-Work}}

While there are related approaches enabling automatic user interface
generation, most of them are not from the domain of process management,
but can be found in the field of end-user programming, including approaches
such as low-code development platforms. Most notably SUPPLE \citep{Gajos2010}
falls into this category, it takes user interface generation and defines
it as constrained optimization problem, with the constraints determined
by external factors, such as usage patterns, screen sizes, or user
accessibility requirements. Basically, a programmer creates a functional
interface specification, which is then used, together with the constraints,
as input for the SUPPLE system. SUPPLE optimizes the user interface
to conform to the constraints, e.g. removing unimportant widgets to
increase text sizes for users with vision problems instead. While
this is interesting research, core ideas of which could be introduced
into the PHILharmonicFlows runtime user interface, it does not offer
the same level of process support for end-users as the concepts presented
in this article.

A related approach more concerned with the support of users in completing
their tasks, than user interface generation, is FLOWer \citep{Berens2005}.
FLOWer is a tool that implements the case handling process support
paradigm. Case handling assumes that workers need access to an entire
case or ``work object'' at any time to complete their tasks. Note
that this is in contrast to traditional activity-centric process support
paradigms, in which users are only presented with the information
they need to complete one specific activity. While FLOWer offers a
sophisticated method for work distribution, i.e., generating to-do
lists for users based on roles and process state, the user interface
itself is not generated directly from the case information. This is
the opposite of an approach like SUPPLE, which is only concerned with
user interface generation.

\citep{Kolb2012} proposes the use of transformation patterns between
process fragments and the user interface. This allows for rapid user
interface development by mapping common patterns found in process
models to reusable form elements. Furthermore, users may re-arrange
the generated user interface, thereby changing the underlying process
model.

\section{Summary and Outlook\label{sec:Summary}}

In summary, the presented contribution sits between these two examples
of related work given in Section \ref{sec:Related-Work}, as it aims
not only to offer process-based workflow support, but also generate
a user interface with forms, navigation, and high level abstractions
for end-users. While there is room for improvement on the design,
or even the usability side of the current PHILharmonicFlows end-user
interface, the importance of creating an actually usable object-aware
process management system can not be stressed enough. This is especially
true when considering the research potential that stems from having
a prototype of an object-aware process management system that can
be deployed in real-world scenarios and utilized, not only by experts,
but also end-users for their daily tasks.

Currently, most research into more advanced topics in the business
process management field, such as process mining, compliance conformity
checking, and the gathering of business intelligence data is done
on the plethora of existing activity-centric process engines and paradigms.
However, a generic user interface for object-aware processes allows
for these research fields to be opened to non-activity-centric processes
as we are now able to collect real-world data by deploying object-aware
processes in the field.

Further note that the log files generated by users when interacting
with the PHILharmonicFlows process engine are very fine-grained, as
the writing of individual attributes is logged separately, due to
the fact that the entire flow of the form logic is determined by the
process engine using the lifecycle processes of the various objects
in an object-aware process model. This opens up opportunities for
research not available in activity-centric process engines, in which
only the data from completed forms attached to activities is communicated
to the server.

Due to the fine-grained logs produced, we will be able to apply approaches
like machine learning to the engine, offering quality of life improvements
such as pre-filling attributes with values based on historical training
data or, even more importantly, optimizing the flow of the generated
user interface by detecting usage patterns, e.g. in navigation and
form usage. The fact that these elements are generated and not hard-coded
in the current user interface prototype allows us to research ways
to improve their generation using the data we collect from our ongoing
studies, real-world deployments, and the aforementioned technologies
such as machine learning. Finally, we intend to apply the lessons
learned in this research to other projects in the field of data-centric
process management.

\subsubsection*{Acknowledgments}

This work is part of the ZAFH Intralogistik, funded by the European
Regional Development Fund and the Ministry of Science, Research and
the Arts of Baden-Württemberg, Germany (F.No. 32-7545.24-17/3/1)

\bibliographystyle{elsarticle-num}
\bibliography{References}

\end{document}